# Supporting Information for the Paper: Optimal Ternary Constant-Composition Codes of Weight Four and Distance Five, IEEE Trans. Inform. Theory, To Appear.

Fei Gao and Gennian Ge

Table 1: Base Codewords of Optimal $(6t+3, 5, [3, 1])_3$-codes

| $t$ | Base Codewords | | | | | |
|---|---|---|---|---|---|---|
| 7 | $\langle 4,3,6,1\rangle$ | $\langle 9,34,4,2\rangle$ | $\langle 3,23,29,2\rangle$ | $\langle 11,16,38,2\rangle$ | $\langle 19,17,10,2\rangle$ | $\langle 33,39,15,2\rangle$ |
| | $\langle 18,6,8,2\rangle$ | $\langle 13,26,5,2\rangle$ | $\langle 31,14,5,1\rangle$ | $\langle 15,26,37,0\rangle$ | $\langle 21,14,30,2\rangle$ | $\langle 33,41,18,1\rangle$ |
| | $\langle 7,0,19,1\rangle$ | $\langle 25,1,28,0\rangle$ | $\langle 7,23,36,0\rangle$ | $\langle 16,10,27,1\rangle$ | $\langle 32,44,29,0\rangle$ | $\langle 35,34,30,1\rangle$ |
| | $\langle 9,22,8,0\rangle$ | $\langle 3,13,17,0\rangle$ | $\langle 8,12,43,1\rangle$ | $\langle 16,24,41,0\rangle$ | | |
| 8 | $\langle 0,12,4,2\rangle$ | $\langle 35,26,9,1\rangle$ | $\langle 18,17,41,2\rangle$ | $\langle 25,27,14,0\rangle$ | $\langle 32,42,11,1\rangle$ | $\langle 36,41,31,1\rangle$ |
| | $\langle 3,4,45,1\rangle$ | $\langle 40,1,12,0\rangle$ | $\langle 18,36,42,0\rangle$ | $\langle 29,35,27,2\rangle$ | $\langle 32,49,14,2\rangle$ | $\langle 39,50,46,2\rangle$ |
| | $\langle 1,26,30,2\rangle$ | $\langle 43,24,7,2\rangle$ | $\langle 19,11,16,2\rangle$ | $\langle 30,46,15,1\rangle$ | $\langle 34,16,14,1\rangle$ | $\langle 44,22,43,0\rangle$ |
| | $\langle 2,21,24,0\rangle$ | $\langle 6,36,20,2\rangle$ | $\langle 22,13,50,1\rangle$ | $\langle 32,39,13,0\rangle$ | $\langle 34,40,47,2\rangle$ | $\langle 49,12,25,1\rangle$ |
| | $\langle 23,8,11,0\rangle$ | | | | | |
| 9 | $\langle 35,9,2,1\rangle$ | $\langle 3,50,56,1\rangle$ | $\langle 10,52,34,2\rangle$ | $\langle 31,44,29,0\rangle$ | $\langle 37,49,17,2\rangle$ | $\langle 40,39,13,0\rangle$ |
| | $\langle 1,15,53,0\rangle$ | $\langle 36,7,30,2\rangle$ | $\langle 11,52,48,0\rangle$ | $\langle 32,43,33,2\rangle$ | $\langle 38,28,56,2\rangle$ | $\langle 41,36,11,1\rangle$ |
| | $\langle 12,24,9,2\rangle$ | $\langle 41,1,22,2\rangle$ | $\langle 23,26,35,2\rangle$ | $\langle 32,55,15,1\rangle$ | $\langle 38,36,54,0\rangle$ | $\langle 49,35,43,0\rangle$ |
| | $\langle 28,37,2,0\rangle$ | $\langle 41,6,27,0\rangle$ | $\langle 26,47,18,0\rangle$ | $\langle 33,20,13,1\rangle$ | $\langle 40,37,42,1\rangle$ | $\langle 51,24,43,1\rangle$ |
| | $\langle 29,28,6,1\rangle$ | $\langle 48,0,16,2\rangle$ | $\langle 27,51,40,2\rangle$ | $\langle 34,38,27,1\rangle$ | | |
| 10 | $\langle 3,1,0,2\rangle$ | $\langle 10,4,18,2\rangle$ | $\langle 32,54,9,0\rangle$ | $\langle 13,31,57,0\rangle$ | $\langle 16,38,57,2\rangle$ | $\langle 19,47,58,0\rangle$ |
| | $\langle 4,5,0,1\rangle$ | $\langle 21,42,8,1\rangle$ | $\langle 7,43,50,1\rangle$ | $\langle 14,40,61,2\rangle$ | $\langle 17,46,41,0\rangle$ | $\langle 20,37,62,1\rangle$ |
| | $\langle 12,6,2,1\rangle$ | $\langle 21,5,12,0\rangle$ | $\langle 8,11,20,2\rangle$ | $\langle 14,47,29,1\rangle$ | $\langle 18,28,57,1\rangle$ | $\langle 21,23,59,2\rangle$ |
| | $\langle 4,34,2,0\rangle$ | $\langle 22,9,35,2\rangle$ | $\langle 10,61,58,1\rangle$ | $\langle 15,23,29,0\rangle$ | $\langle 19,30,42,2\rangle$ | $\langle 24,41,49,1\rangle$ |
| | $\langle 6,1,36,0\rangle$ | $\langle 3,14,10,0\rangle$ | $\langle 12,43,27,2\rangle$ | $\langle 16,35,25,1\rangle$ | $\langle 19,39,59,1\rangle$ | $\langle 25,11,42,0\rangle$ |
| | $\langle 6,5,33,2\rangle$ | | | | | |
| 11 | $\langle 12,2,6,1\rangle$ | $\langle 19,2,56,0\rangle$ | $\langle 14,67,52,1\rangle$ | $\langle 23,66,10,0\rangle$ | $\langle 30,15,46,0\rangle$ | $\langle 36,16,56,1\rangle$ |
| | $\langle 17,5,8,1\rangle$ | $\langle 27,59,8,0\rangle$ | $\langle 16,61,41,0\rangle$ | $\langle 24,41,49,1\rangle$ | $\langle 30,25,14,2\rangle$ | $\langle 36,28,35,0\rangle$ |
| | $\langle 7,1,34,0\rangle$ | $\langle 6,20,14,0\rangle$ | $\langle 19,40,42,1\rangle$ | $\langle 25,51,39,0\rangle$ | $\langle 31,26,28,1\rangle$ | $\langle 38,33,37,0\rangle$ |
| | $\langle 0,60,62,1\rangle$ | $\langle 9,43,31,2\rangle$ | $\langle 20,51,34,1\rangle$ | $\langle 25,68,47,1\rangle$ | $\langle 32,28,62,2\rangle$ | $\langle 40,17,50,0\rangle$ |
| | $\langle 13,32,4,0\rangle$ | $\langle 11,24,53,0\rangle$ | $\langle 21,22,40,2\rangle$ | $\langle 26,15,50,2\rangle$ | $\langle 33,61,22,1\rangle$ | $\langle 42,66,20,2\rangle$ |
| | $\langle 18,39,0,2\rangle$ | $\langle 12,52,45,0\rangle$ | $\langle 21,58,48,0\rangle$ | $\langle 29,54,57,0\rangle$ | | |
| 13 | $\langle 0,5,4,1\rangle$ | $\langle 3,11,78,1\rangle$ | $\langle 9,26,80,2\rangle$ | $\langle 13,66,34,1\rangle$ | $\langle 17,77,37,2\rangle$ | $\langle 20,43,77,1\rangle$ |
| | $\langle 1,0,3,2\rangle$ | $\langle 4,10,46,2\rangle$ | $\langle 9,40,48,1\rangle$ | $\langle 14,29,47,1\rangle$ | $\langle 18,47,77,0\rangle$ | $\langle 20,64,48,0\rangle$ |
| | $\langle 1,6,8,0\rangle$ | $\langle 5,21,12,0\rangle$ | $\langle 10,25,56,1\rangle$ | $\langle 14,33,53,2\rangle$ | $\langle 18,56,64,2\rangle$ | $\langle 22,38,76,2\rangle$ |
| | $\langle 2,4,7,0\rangle$ | $\langle 7,19,30,2\rangle$ | $\langle 11,20,23,2\rangle$ | $\langle 15,36,59,0\rangle$ | $\langle 18,59,65,1\rangle$ | $\langle 22,46,71,1\rangle$ |
| | $\langle 2,6,51,1\rangle$ | $\langle 7,79,26,1\rangle$ | $\langle 12,24,80,1\rangle$ | $\langle 15,37,63,1\rangle$ | $\langle 19,41,70,0\rangle$ | $\langle 23,37,54,0\rangle$ |
| | $\langle 5,6,16,2\rangle$ | $\langle 8,34,44,2\rangle$ | $\langle 13,51,33,0\rangle$ | $\langle 15,39,69,2\rangle$ | $\langle 19,60,75,1\rangle$ | $\langle 30,67,49,0\rangle$ |
| | $\langle 10,3,14,0\rangle$ | $\langle 9,22,35,0\rangle$ | $\langle 13,61,27,2\rangle$ | $\langle 17,63,34,0\rangle$ | | |
| 14 | $\langle 1, 62, 47, 0\rangle$ | $\langle 5, 43, 27, 0\rangle$ | $\langle 14, 72, 48, 2\rangle$ | $\langle 29, 39, 73, 2\rangle$ | $\langle 41, 30, 33, 1\rangle$ | $\langle 67, 60, 10, 1\rangle$ |
| | $\langle 15, 21, 6, 2\rangle$ | $\langle 74, 3, 17, 1\rangle$ | $\langle 15, 11, 19, 0\rangle$ | $\langle 30, 17, 70, 2\rangle$ | $\langle 42, 54, 67, 0\rangle$ | $\langle 69, 12, 33, 2\rangle$ |
| | $\langle 20, 14, 7, 0\rangle$ | $\langle 77, 78, 9, 1\rangle$ | $\langle 16, 25, 83, 2\rangle$ | $\langle 30, 69, 40, 0\rangle$ | $\langle 42, 85, 64, 1\rangle$ | $\langle 74, 70, 18, 0\rangle$ |
| | $\langle 22, 63, 4, 2\rangle$ | $\langle 8, 59, 20, 1\rangle$ | $\langle 16, 86, 36, 1\rangle$ | $\langle 31, 32, 82, 2\rangle$ | $\langle 43, 59, 49, 2\rangle$ | $\langle 78, 46, 35, 2\rangle$ |
| | $\langle 3, 52, 67, 2\rangle$ | $\langle 85, 9, 50, 2\rangle$ | $\langle 24, 56, 51, 1\rangle$ | $\langle 34, 11, 76, 1\rangle$ | $\langle 45, 43, 40, 1\rangle$ | $\langle 79, 31, 48, 1\rangle$ |
| | $\langle 32, 6, 60, 0\rangle$ | $\langle 9, 82, 28, 0\rangle$ | $\langle 27, 29, 50, 1\rangle$ | $\langle 37, 13, 32, 1\rangle$ | $\langle 47, 71, 74, 2\rangle$ | $\langle 80, 26, 82, 1\rangle$ |
| | $\langle 38, 3, 45, 0\rangle$ | $\langle 14, 23, 63, 1\rangle$ | $\langle 28, 54, 55, 1\rangle$ | $\langle 37, 65, 25, 0\rangle$ | $\langle 66, 83, 58, 0\rangle$ | $\langle 86, 13, 68, 2\rangle$ |
| | $\langle 47, 5, 72, 1\rangle$ | | | | | |
| 15 | $\langle 13, 6, 8, 0\rangle$ | $\langle 39, 85, 7, 0\rangle$ | $\langle 92, 43, 3, 2\rangle$ | $\langle 24, 52, 73, 0\rangle$ | $\langle 41, 76, 26, 1\rangle$ | $\langle 76, 35, 59, 0\rangle$ |
| | $\langle 9, 30, 1, 2\rangle$ | $\langle 56, 28, 1, 0\rangle$ | $\langle 12, 83, 91, 1\rangle$ | $\langle 29, 88, 76, 2\rangle$ | $\langle 44, 65, 75, 2\rangle$ | $\langle 77, 59, 78, 1\rangle$ |
| | $\langle 10, 4, 66, 1\rangle$ | $\langle 6, 59, 19, 2\rangle$ | $\langle 17, 24, 63, 2\rangle$ | $\langle 31, 13, 51, 2\rangle$ | $\langle 44, 67, 37, 0\rangle$ | $\langle 81, 72, 46, 1\rangle$ |
| | $\langle 11, 48, 7, 2\rangle$ | $\langle 60, 3, 30, 0\rangle$ | $\langle 17, 28, 29, 1\rangle$ | $\langle 36, 23, 12, 0\rangle$ | $\langle 55, 16, 39, 2\rangle$ | $\langle 82, 53, 11, 1\rangle$ |
| | $\langle 2, 27, 78, 0\rangle$ | $\langle 62, 35, 5, 2\rangle$ | $\langle 17, 50, 45, 0\rangle$ | $\langle 37, 56, 46, 2\rangle$ | $\langle 63, 71, 30, 1\rangle$ | $\langle 84, 20, 69, 2\rangle$ |
| | $\langle 2, 69, 92, 1\rangle$ | $\langle 7, 49, 52, 1\rangle$ | $\langle 21, 64, 66, 0\rangle$ | $\langle 37, 74, 42, 1\rangle$ | $\langle 66, 49, 80, 2\rangle$ | $\langle 88, 74, 68, 0\rangle$ |
| | $\langle 29, 63, 4, 0\rangle$ | $\langle 87, 71, 4, 2\rangle$ | $\langle 22, 33, 83, 2\rangle$ | $\langle 40, 36, 39, 1\rangle$ | $\langle 68, 23, 77, 2\rangle$ | $\langle 89, 73, 54, 2\rangle$ |
| | $\langle 31, 6, 44, 1\rangle$ | $\langle 9, 84, 90, 1\rangle$ | $\langle 22, 91, 55, 0\rangle$ | $\langle 40, 38, 18, 0\rangle$ | | |
| 16 | $\langle 9, 5, 3, 1\rangle$ | $\langle 39, 5, 57, 2\rangle$ | $\langle 19, 59, 39, 0\rangle$ | $\langle 34, 97, 57, 0\rangle$ | $\langle 62, 63, 31, 0\rangle$ | $\langle 78, 75, 42, 1\rangle$ |
| | $\langle 1, 11, 14, 0\rangle$ | $\langle 52, 18, 6, 0\rangle$ | $\langle 21, 51, 30, 0\rangle$ | $\langle 35, 83, 64, 0\rangle$ | $\langle 64, 42, 53, 2\rangle$ | $\langle 79, 72, 65, 0\rangle$ |
| | $\langle 14, 51, 0, 2\rangle$ | $\langle 56, 4, 31, 1\rangle$ | $\langle 21, 74, 45, 1\rangle$ | $\langle 49, 73, 28, 1\rangle$ | $\langle 65, 89, 81, 1\rangle$ | $\langle 80, 97, 95, 1\rangle$ |
| | $\langle 15, 7, 10, 0\rangle$ | $\langle 6, 19, 66, 2\rangle$ | $\langle 24, 80, 79, 2\rangle$ | $\langle 50, 91, 29, 2\rangle$ | $\langle 66, 82, 94, 0\rangle$ | $\langle 81, 96, 41, 2\rangle$ |
| | $\langle 2, 55, 22, 0\rangle$ | $\langle 72, 8, 50, 1\rangle$ | $\langle 25, 86, 36, 0\rangle$ | $\langle 53, 44, 71, 1\rangle$ | $\langle 68, 19, 34, 1\rangle$ | $\langle 87, 97, 15, 2\rangle$ |
| | $\langle 26, 9, 80, 0\rangle$ | $\langle 10, 39, 58, 1\rangle$ | $\langle 28, 58, 50, 0\rangle$ | $\langle 57, 98, 55, 1\rangle$ | $\langle 71, 16, 32, 0\rangle$ | $\langle 90, 52, 59, 1\rangle$ |
| | $\langle 27, 8, 84, 0\rangle$ | $\langle 12, 86, 13, 1\rangle$ | $\langle 29, 23, 96, 1\rangle$ | $\langle 60, 64, 15, 1\rangle$ | $\langle 73, 34, 16, 2\rangle$ | $\langle 92, 23, 56, 2\rangle$ |
| | $\langle 3, 41, 46, 0\rangle$ | $\langle 17, 49, 12, 0\rangle$ | $\langle 30, 16, 20, 1\rangle$ | $\langle 61, 38, 26, 1\rangle$ | $\langle 76, 45, 70, 0\rangle$ | $\langle 94, 85, 21, 2\rangle$ |
| | $\langle 35, 7, 48, 1\rangle$ | | | | | |
| 18 | $\langle 0, 68, 9, 1\rangle$ | $\langle 8, 76, 97, 0\rangle$ | $\langle 16, 20, 93, 1\rangle$ | $\langle 6, 55, 102, 2\rangle$ | $\langle 83, 73, 64, 1\rangle$ | $\langle 100, 92, 46, 2\rangle$ |
| | $\langle 18, 8, 11, 2\rangle$ | $\langle 99, 61, 0, 2\rangle$ | $\langle 16, 22, 36, 2\rangle$ | $\langle 67, 48, 22, 0\rangle$ | $\langle 90, 77, 32, 1\rangle$ | $\langle 101, 27, 23, 1\rangle$ |
| | $\langle 2, 98, 99, 1\rangle$ | $\langle 10, 68, 51, 2\rangle$ | $\langle 17, 36, 65, 0\rangle$ | $\langle 7, 86, 103, 0\rangle$ | $\langle 90, 82, 24, 2\rangle$ | $\langle 103, 92, 31, 1\rangle$ |
| | $\langle 23, 52, 5, 0\rangle$ | $\langle 105, 84, 4, 1\rangle$ | $\langle 25, 37, 21, 0\rangle$ | $\langle 72, 85, 18, 0\rangle$ | $\langle 94, 42, 12, 2\rangle$ | $\langle 104, 54, 97, 1\rangle$ |
| | $\langle 39, 14, 3, 1\rangle$ | $\langle 11, 94, 71, 0\rangle$ | $\langle 40, 43, 38, 2\rangle$ | $\langle 75, 15, 59, 0\rangle$ | $\langle 97, 35, 59, 2\rangle$ | $\langle 106, 37, 30, 2\rangle$ |
| | $\langle 4, 67, 85, 2\rangle$ | $\langle 12, 60, 95, 1\rangle$ | $\langle 45, 43, 42, 1\rangle$ | $\langle 78, 54, 32, 2\rangle$ | $\langle 98, 14, 26, 2\rangle$ | $\langle 107, 53, 30, 0\rangle$ |
| | $\langle 4, 82, 56, 0\rangle$ | $\langle 14, 109, 9, 0\rangle$ | $\langle 46, 11, 47, 1\rangle$ | $\langle 78, 8, 110, 1\rangle$ | $\langle 99, 60, 66, 0\rangle$ | $\langle 108, 34, 66, 2\rangle$ |
| | $\langle 6, 68, 28, 0\rangle$ | $\langle 15, 10, 33, 1\rangle$ | $\langle 52, 38, 69, 1\rangle$ | $\langle 80, 13, 50, 2\rangle$ | $\langle 10, 101, 81, 0\rangle$ | $\langle 109, 29, 84, 2\rangle$ |
| | $\langle 7, 62, 20, 2\rangle$ | $\langle 15, 23, 17, 2\rangle$ | $\langle 57, 85, 30, 1\rangle$ | $\langle 83, 12, 58, 0\rangle$ | $\langle 100, 13, 73, 0\rangle$ | $\langle 110, 89, 63, 0\rangle$ |
| | $\langle 7, 82, 41, 1\rangle$ | | | | | |
| 19 | $\langle 6, 8, 1, 0\rangle$ | $\langle 10, 68, 79, 2\rangle$ | $\langle 25, 98, 88, 2\rangle$ | $\langle 34, 91, 29, 2\rangle$ | $\langle 21, 10, 115, 1\rangle$ | $\langle 32, 104, 67, 2\rangle$ |
| | $\langle 9, 1, 7, 2\rangle$ | $\langle 11, 67, 64, 1\rangle$ | $\langle 27, 77, 99, 0\rangle$ | $\langle 35, 42, 51, 1\rangle$ | $\langle 21, 112, 94, 2\rangle$ | $\langle 32, 82, 101, 1\rangle$ |
| | $\langle 0, 5, 93, 2\rangle$ | $\langle 14, 46, 29, 1\rangle$ | $\langle 28, 43, 59, 1\rangle$ | $\langle 35, 61, 95, 2\rangle$ | $\langle 22, 84, 111, 0\rangle$ | $\langle 37, 114, 17, 2\rangle$ |
| | $\langle 10, 14, 3, 0\rangle$ | $\langle 15, 19, 94, 1\rangle$ | $\langle 28, 71, 49, 0\rangle$ | $\langle 35, 7, 101, 0\rangle$ | $\langle 23, 107, 53, 1\rangle$ | $\langle 38, 115, 76, 2\rangle$ |
| | $\langle 18, 34, 3, 1\rangle$ | $\langle 18, 82, 55, 2\rangle$ | $\langle 30, 55, 47, 0\rangle$ | $\langle 36, 32, 42, 0\rangle$ | $\langle 23, 108, 50, 2\rangle$ | $\langle 39, 75, 113, 2\rangle$ |
| | $\langle 22, 52, 0, 1\rangle$ | $\langle 18, 83, 85, 0\rangle$ | $\langle 30, 92, 16, 2\rangle$ | $\langle 37, 24, 50, 1\rangle$ | $\langle 25, 113, 61, 0\rangle$ | $\langle 40, 75, 110, 1\rangle$ |
| | $\langle 33, 16, 2, 0\rangle$ | $\langle 19, 78, 48, 2\rangle$ | $\langle 31, 63, 82, 0\rangle$ | $\langle 38, 13, 62, 1\rangle$ | $\langle 26, 104, 62, 0\rangle$ | $\langle 5, 116, 108, 0\rangle$ |
| | $\langle 37, 78, 4, 0\rangle$ | $\langle 20, 38, 57, 0\rangle$ | $\langle 33, 80, 12, 2\rangle$ | $\langle 57, 44, 69, 1\rangle$ | $\langle 27, 83, 111, 1\rangle$ | $\langle 8, 113, 116, 1\rangle$ |
| | $\langle 6, 55, 54, 1\rangle$ | $\langle 20, 66, 84, 2\rangle$ | $\langle 33, 90, 36, 1\rangle$ | $\langle 13, 96, 106, 0\rangle$ | $\langle 30, 71, 105, 1\rangle$ | $\langle 24, 102, 101, 2\rangle$ |
| | $\langle 9, 70, 79, 1\rangle$ | $\langle 21, 67, 87, 0\rangle$ | $\langle 34, 80, 79, 0\rangle$ | $\langle 20, 114, 41, 1\rangle$ | | |
| 20 | $\langle 66,9,7,1\rangle$ | $\langle 25,91,22,1\rangle$ | $\langle 52,46,60,1\rangle$ | $\langle 74,81,90,1\rangle$ | $\langle 38,109,41,1\rangle$ | $\langle 65,19,103,1\rangle$ |
| | $\langle 19,18,0,2\rangle$ | $\langle 27,102,7,0\rangle$ | $\langle 54,97,71,0\rangle$ | $\langle 75,63,85,0\rangle$ | $\langle 42,118,12,0\rangle$ | $\langle 66,101,31,2\rangle$ |
| | $\langle 29,88,8,2\rangle$ | $\langle 27,82,64,1\rangle$ | $\langle 58,98,16,2\rangle$ | $\langle 80,2,112,1\rangle$ | $\langle 46,69,108,0\rangle$ | $\langle 68,115,59,0\rangle$ |
| | $\langle 30,3,45,0\rangle$ | $\langle 30,80,81,2\rangle$ | $\langle 61,13,28,0\rangle$ | $\langle 83,51,20,0\rangle$ | $\langle 48,117,10,1\rangle$ | $\langle 72,36,110,1\rangle$ |
| | $\langle 41,10,5,2\rangle$ | $\langle 35,81,57,0\rangle$ | $\langle 61,34,87,2\rangle$ | $\langle 90,95,94,0\rangle$ | $\langle 50,120,26,2\rangle$ | $\langle 73,109,15,2\rangle$ |
| | $\langle 42,5,23,1\rangle$ | $\langle 35,99,65,2\rangle$ | $\langle 62,73,95,1\rangle$ | $\langle 32,118,40,1\rangle$ | $\langle 52,113,86,2\rangle$ | $\langle 77,102,89,2\rangle$ |
| | $\langle 50,1,25,0\rangle$ | $\langle 39,18,62,0\rangle$ | $\langle 62,76,51,2\rangle$ | $\langle 32,75,116,2\rangle$ | $\langle 53,104,25,2\rangle$ | $\langle 78,111,22,0\rangle$ |
| | $\langle 14,20,43,2\rangle$ | $\langle 46,39,11,2\rangle$ | $\langle 64,74,55,0\rangle$ | $\langle 33,112,95,2\rangle$ | $\langle 60,19,112,0\rangle$ | $\langle 39,122,107,1\rangle$ |
| | $\langle 16,89,14,1\rangle$ | $\langle 48,38,42,2\rangle$ | $\langle 71,15,29,1\rangle$ | $\langle 36,44,101,0\rangle$ | $\langle 60,57,106,2\rangle$ | $\langle 72,119,103,0\rangle$ |
| | $\langle 2,56,121,0\rangle$ | $\langle 50,30,70,1\rangle$ | $\langle 74,67,72,2\rangle$ | $\langle 37,100,49,0\rangle$ | $\langle 63,108,11,1\rangle$ | $\langle 88,120,109,0\rangle$ |
| | $\langle 24,87,37,1\rangle$ | | | | | |
| 21 | $\langle 1,6,31,2\rangle$ | $\langle 23,19,71,0\rangle$ | $\langle 48,38,89,2\rangle$ | $\langle 10,27,127,0\rangle$ | $\langle 122,65,54,2\rangle$ | $\langle 81,104,12,2\rangle$ |
| | $\langle 15,3,6,1\rangle$ | $\langle 24,30,37,0\rangle$ | $\langle 53,25,21,0\rangle$ | $\langle 105,20,84,2\rangle$ | $\langle 124,41,83,1\rangle$ | $\langle 86,110,81,1\rangle$ |
| | $\langle 11,5,36,2\rangle$ | $\langle 24,32,61,2\rangle$ | $\langle 62,29,92,1\rangle$ | $\langle 108,15,81,0\rangle$ | $\langle 127,83,68,2\rangle$ | $\langle 94,67,120,1\rangle$ |
| | $\langle 47,2,93,1\rangle$ | $\langle 28,35,34,0\rangle$ | $\langle 64,85,25,2\rangle$ | $\langle 108,28,95,2\rangle$ | $\langle 128,42,34,1\rangle$ | $\langle 99,117,77,0\rangle$ |
| | $\langle 5,23,73,1\rangle$ | $\langle 30,78,97,2\rangle$ | $\langle 75,114,7,1\rangle$ | $\langle 109,14,61,1\rangle$ | $\langle 16,52,117,2\rangle$ | $\langle 105,103,31,1\rangle$ |
| | $\langle 5,39,85,0\rangle$ | $\langle 32,36,92,0\rangle$ | $\langle 8,120,62,0\rangle$ | $\langle 112,60,16,1\rangle$ | $\langle 60,91,111,0\rangle$ | $\langle 114,11,113,0\rangle$ |
| | $\langle 114,62,9,2\rangle$ | $\langle 37,3,113,2\rangle$ | $\langle 84,49,46,1\rangle$ | $\langle 115,52,99,1\rangle$ | $\langle 72,57,102,2\rangle$ | $\langle 116,100,85,1\rangle$ |
| | $\langle 119,6,79,0\rangle$ | $\langle 38,26,13,0\rangle$ | $\langle 87,1,107,0\rangle$ | $\langle 117,28,50,1\rangle$ | $\langle 73,64,115,0\rangle$ | $\langle 12,100,118,0\rangle$ |
| | $\langle 12,82,71,1\rangle$ | $\langle 43,41,80,2\rangle$ | $\langle 96,106,9,0\rangle$ | $\langle 119,10,64,1\rangle$ | $\langle 74,110,60,2\rangle$ | $\langle 90,123,125,0\rangle$ |
| | $\langle 128,4,54,0\rangle$ | $\langle 46,65,56,0\rangle$ | $\langle 96,14,42,2\rangle$ | $\langle 119,22,46,2\rangle$ | $\langle 80,101,37,1\rangle$ | $\langle 95,102,103,0\rangle$ |
| | $\langle 19,33,90,1\rangle$ | $\langle 47,61,44,0\rangle$ | $\langle 10,21,101,2\rangle$ | $\langle 121,76,50,0\rangle$ | | |

Table 2: Base Codewords of $[3,1]$-GDC$(5)$s of Type $2^{3u}4^1$

| $u$ | Base Codewords | | | | | |
|---|---|---|---|---|---|---|
| 3 | $\langle 0,2,8,16\rangle$ | $\langle 0,11,14,17\rangle$ | $\langle \infty_1,0,7,4\rangle$ | $\langle \infty_0,0,1,11\rangle$ | $\langle \infty_2,0,13,15\rangle$ | $\langle \infty_3,0,17,12\rangle$ |
| | $\langle 1,5,11,4\rangle$ | $\langle 1,14,17,15\rangle$ | | | | |
| 4 | $\langle 0,1,2,19\rangle$ | $\langle 0,4,10,13\rangle$ | $\langle 1,4,23,15\rangle$ | $\langle \infty_0,0,3,1\rangle$ | $\langle \infty_2,0,11,20\rangle$ | $\langle \infty_3,0,17,21\rangle$ |
| | $\langle 1,7,21,2\rangle$ | $\langle 0,8,15,14\rangle$ | $\langle 1,9,20,12\rangle$ | $\langle \infty_1,0,9,2\rangle$ | | |
| 5 | $\langle 0,4,9,5\rangle$ | $\langle 1,6,7,12\rangle$ | $\langle 0,3,23,26\rangle$ | $\langle 1,5,13,25\rangle$ | $\langle \infty_0,0,7,8\rangle$ | $\langle \infty_2,0,17,16\rangle$ |
| | $\langle 0,8,24,3\rangle$ | $\langle 0,18,20,2\rangle$ | $\langle 1,2,15,23\rangle$ | $\langle 1,10,29,20\rangle$ | $\langle \infty_1,0,11,13\rangle$ | $\langle \infty_3,0,27,24\rangle$ |
| 6 | $\langle 0,8,30,2\rangle$ | $\langle 0,20,27,1\rangle$ | $\langle 1,5,22,35\rangle$ | $\langle 0,12,21,23\rangle$ | $\langle \infty_0,0,5,21\rangle$ | $\langle \infty_2,0,29,6\rangle$ |
| | $\langle 1,4,6,18\rangle$ | $\langle 0,4,17,20\rangle$ | $\langle 0,10,11,19\rangle$ | $\langle 1,11,13,22\rangle$ | $\langle \infty_1,0,25,3\rangle$ | $\langle \infty_3,0,35,24\rangle$ |
| | $\langle 1,9,15,6\rangle$ | $\langle 1,14,17,0\rangle$ | | | | |
| 7 | $\langle 0,4,31,9\rangle$ | $\langle 0,1,40,29\rangle$ | $\langle 0,5,30,41\rangle$ | $\langle 1,5,18,15\rangle$ | $\langle \infty_0,0,9,2\rangle$ | $\langle \infty_2,0,37,10\rangle$ |
| | $\langle 0,6,20,1\rangle$ | $\langle 0,10,34,6\rangle$ | $\langle 1,2,35,10\rangle$ | $\langle 1,15,17,33\rangle$ | $\langle \infty_1,0,15,3\rangle$ | $\langle \infty_3,0,39,16\rangle$ |
| | $\langle 1,8,19,2\rangle$ | $\langle 0,16,23,4\rangle$ | $\langle 1,24,37,0\rangle$ | $\langle 1,21,33,13\rangle$ | | |
| 8 | $\langle 0,3,30,4\rangle$ | $\langle 0,6,14,23\rangle$ | $\langle 1,4,29,39\rangle$ | $\langle 0,27,36,31\rangle$ | $\langle 1,18,20,47\rangle$ | $\langle \infty_1,0,13,2\rangle$ |
| | $\langle 0,1,38,46\rangle$ | $\langle 0,7,22,41\rangle$ | $\langle 1,5,30,45\rangle$ | $\langle 0,28,44,10\rangle$ | $\langle 1,34,43,28\rangle$ | $\langle \infty_2,0,37,1\rangle$ |
| | $\langle 0,17,47,5\rangle$ | $\langle 1,3,17,26\rangle$ | $\langle 1,9,14,40\rangle$ | $\langle 1,13,23,18\rangle$ | $\langle \infty_0,0,5,7\rangle$ | $\langle \infty_3,0,41,6\rangle$ |
| 9 | $\langle 0,7,28,2\rangle$ | $\langle 0,13,42,6\rangle$ | $\langle 1,5,38,20\rangle$ | $\langle 0,16,34,37\rangle$ | $\langle 1,16,31,29\rangle$ | $\langle \infty_1,0,11,7\rangle$ |
| | $\langle 1,3,6,49\rangle$ | $\langle 0,5,50,30\rangle$ | $\langle 1,9,18,44\rangle$ | $\langle 0,22,24,10\rangle$ | $\langle 1,17,37,43\rangle$ | $\langle \infty_2,0,31,1\rangle$ |
| | $\langle 1,8,49,0\rangle$ | $\langle 0,6,46,14\rangle$ | $\langle 0,10,29,33\rangle$ | $\langle 1,13,27,45\rangle$ | $\langle \infty_0,0,3,4\rangle$ | $\langle \infty_3,0,35,15\rangle$ |
| | $\langle 0,1,23,32\rangle$ | $\langle 1,2,45,11\rangle$ | | | | |
| 10 | $\langle 0,2,6,39\rangle$ | $\langle 1,41,45,2\rangle$ | $\langle 0,19,37,18\rangle$ | $\langle 1,12,36,35\rangle$ | $\langle 1,29,30,14\rangle$ | $\langle \infty_1,0,9,2\rangle$ |
| | $\langle 1,7,9,47\rangle$ | $\langle 0,10,57,55\rangle$ | $\langle 0,20,52,11\rangle$ | $\langle 1,15,27,19\rangle$ | $\langle 1,34,51,26\rangle$ | $\langle \infty_2,0,29,5\rangle$ |
| | $\langle 0,22,35,3\rangle$ | $\langle 0,11,18,25\rangle$ | $\langle 0,23,44,13\rangle$ | $\langle 1,16,37,52\rangle$ | $\langle \infty_0,0,1,4\rangle$ | $\langle \infty_3,0,43,6\rangle$ |
| | $\langle 0,7,26,12\rangle$ | $\langle 0,12,15,54\rangle$ | $\langle 1,10,56,34\rangle$ | $\langle 1,23,28,43\rangle$ | | |
| 11 | $\langle 0,2,3,65\rangle$ | $\langle 1,4,65,18\rangle$ | $\langle 0,17,36,11\rangle$ | $\langle 0,40,46,61\rangle$ | $\langle 1,22,29,11\rangle$ | $\langle \infty_0,0,11,3\rangle$ |
| | $\langle 0,5,62,8\rangle$ | $\langle 1,45,55,2\rangle$ | $\langle 0,19,42,58\rangle$ | $\langle 1,10,59,47\rangle$ | $\langle 1,30,53,57\rangle$ | $\langle \infty_1,0,29,1\rangle$ |
| | $\langle 0,16,38,7\rangle$ | $\langle 1,8,61,46\rangle$ | $\langle 0,21,56,46\rangle$ | $\langle 1,19,35,37\rangle$ | $\langle 1,32,40,64\rangle$ | $\langle \infty_2,0,41,4\rangle$ |
| | $\langle 0,18,32,5\rangle$ | $\langle 0,13,54,22\rangle$ | $\langle 0,39,65,26\rangle$ | $\langle 1,21,63,56\rangle$ | $\langle 1,37,52,58\rangle$ | $\langle \infty_3,0,55,9\rangle$ |
| 13 | $\langle 0,8,64,7\rangle$ | $\langle 1,22,75,8\rangle$ | $\langle 0,17,72,19\rangle$ | $\langle 0,49,67,59\rangle$ | $\langle 1,34,77,70\rangle$ | $\langle \infty_0,0,7,3\rangle$ |
| | $\langle 1,6,47,7\rangle$ | $\langle 1,46,55,9\rangle$ | $\langle 0,19,63,50\rangle$ | $\langle 0,65,77,60\rangle$ | $\langle 1,39,42,54\rangle$ | $\langle \infty_1,0,25,5\rangle$ |
| | $\langle 0,12,30,9\rangle$ | $\langle 1,68,69,0\rangle$ | $\langle 0,26,54,42\rangle$ | $\langle 1,10,71,14\rangle$ | $\langle 1,49,63,18\rangle$ | $\langle \infty_2,0,31,13\rangle$ |
| | $\langle 0,13,55,2\rangle$ | $\langle 1,8,76,60\rangle$ | $\langle 0,27,76,45\rangle$ | $\langle 1,21,64,41\rangle$ | $\langle 1,51,73,45\rangle$ | $\langle \infty_3,0,59,11\rangle$ |
| | $\langle 0,4,38,30\rangle$ | $\langle 0,16,58,51\rangle$ | $\langle 0,46,51,54\rangle$ | $\langle 1,27,58,13\rangle$ | | |
| 14 | $\langle 0,2,3,34\rangle$ | $\langle 1,21,30,0\rangle$ | $\langle 1,9,31,24\rangle$ | $\langle 0,22,60,20\rangle$ | $\langle 1,11,51,65\rangle$ | $\langle 1,76,83,31\rangle$ |
| | $\langle 1,4,69,7\rangle$ | $\langle 1,40,56,8\rangle$ | $\langle 0,10,51,59\rangle$ | $\langle 0,27,58,53\rangle$ | $\langle 1,12,48,63\rangle$ | $\langle \infty_0,0,35,2\rangle$ |
| | $\langle 0,18,23,7\rangle$ | $\langle 1,6,70,10\rangle$ | $\langle 0,11,47,63\rangle$ | $\langle 0,32,71,67\rangle$ | $\langle 1,13,71,62\rangle$ | $\langle \infty_2,0,69,6\rangle$ |
| | $\langle 0,4,54,13\rangle$ | $\langle 1,7,53,19\rangle$ | $\langle 0,13,72,48\rangle$ | $\langle 0,43,61,46\rangle$ | $\langle 1,25,52,45\rangle$ | $\langle \infty_1,0,67,10\rangle$ |
| | $\langle 0,6,76,33\rangle$ | $\langle 1,8,36,74\rangle$ | $\langle 0,17,21,22\rangle$ | $\langle 0,44,63,65\rangle$ | $\langle 1,29,54,46\rangle$ | $\langle \infty_3,0,83,12\rangle$ |
| 17 | $\langle 0,4,7,41\rangle$ | $\langle 1,12,48,0\rangle$ | $\langle 0,26,78,94\rangle$ | $\langle 0,46,54,21\rangle$ | $\langle 1,38,87,73\rangle$ | $\langle 0,28,57,100\rangle$ |
| | $\langle 1,2,8,26\rangle$ | $\langle 1,47,53,6\rangle$ | $\langle 0,30,90,75\rangle$ | $\langle 1,20,40,49\rangle$ | $\langle 1,39,63,80\rangle$ | $\langle 1,78,99,100\rangle$ |
| | $\langle 0,1,44,58\rangle$ | $\langle 1,77,89,8\rangle$ | $\langle 0,31,40,59\rangle$ | $\langle 1,22,92,60\rangle$ | $\langle 1,4,101,53\rangle$ | $\langle \infty_0,0,27,3\rangle$ |
| | $\langle 0,17,86,1\rangle$ | $\langle 0,10,85,91\rangle$ | $\langle 0,35,69,44\rangle$ | $\langle 1,26,60,19\rangle$ | $\langle 1,50,88,83\rangle$ | $\langle \infty_1,0,39,2\rangle$ |
| | $\langle 0,2,89,78\rangle$ | $\langle 0,14,23,99\rangle$ | $\langle 0,41,73,39\rangle$ | $\langle 1,31,59,13\rangle$ | $\langle 1,61,81,21\rangle$ | $\langle \infty_2,0,47,5\rangle$ |
| | $\langle 0,61,71,8\rangle$ | $\langle 0,18,37,73\rangle$ | $\langle 0,45,80,67\rangle$ | $\langle 1,37,85,28\rangle$ | $\langle 1,90,95,76\rangle$ | $\langle \infty_3,0,79,6\rangle$ |
| 18 | $\langle 0,1,16,24\rangle$ | $\langle 0,20,85,79\rangle$ | $\langle 0,78,87,98\rangle$ | $\langle 1,4,56,107\rangle$ | $\langle 1,59,75,50\rangle$ | $\langle 1,98,102,22\rangle$ |
| | $\langle 0,3,22,16\rangle$ | $\langle 0,24,64,37\rangle$ | $\langle 1,14,104,0\rangle$ | $\langle 1,40,73,88\rangle$ | $\langle 1,85,94,43\rangle$ | $\langle 1,101,103,15\rangle$ |
| | $\langle 0,6,14,69\rangle$ | $\langle 0,25,26,88\rangle$ | $\langle 1,21,43,30\rangle$ | $\langle 1,42,105,9\rangle$ | $\langle 0,12,103,22\rangle$ | $\langle \infty_0,0,39,2\rangle$ |
| | $\langle 1,36,82,5\rangle$ | $\langle 0,28,51,92\rangle$ | $\langle 1,26,28,51\rangle$ | $\langle 1,47,57,35\rangle$ | $\langle 0,34,70,104\rangle$ | $\langle \infty_1,0,43,3\rangle$ |
| | $\langle 0,10,58,99\rangle$ | $\langle 0,35,66,67\rangle$ | $\langle 1,29,97,32\rangle$ | $\langle 1,49,96,17\rangle$ | $\langle 0,57,101,93\rangle$ | $\langle \infty_2,0,45,7\rangle$ |
| | $\langle 0,17,31,68\rangle$ | $\langle 0,37,75,65\rangle$ | $\langle 1,33,62,81\rangle$ | $\langle 1,54,83,93\rangle$ | $\langle 1,12,88,102\rangle$ | $\langle \infty_3,0,71,9\rangle$ |
| | $\langle 0,19,49,96\rangle$ | $\langle 0,41,59,97\rangle$ | | | | |
| 22 | $\langle 1,2,5,77\rangle$ | $\langle 0,5,128,24\rangle$ | $\langle 1,16,31,27\rangle$ | $\langle 0,13,120,42\rangle$ | $\langle 0,82,106,74\rangle$ | $\langle 1,99,117,38\rangle$ |
| | $\langle 0,7,59,27\rangle$ | $\langle 0,55,79,46\rangle$ | $\langle 1,18,34,32\rangle$ | $\langle 0,22,39,127\rangle$ | $\langle 1,20,127,33\rangle$ | $\langle 1,84,125,115\rangle$ |
| | $\langle 1,3,63,99\rangle$ | $\langle 0,58,94,61\rangle$ | $\langle 1,29,58,75\rangle$ | $\langle 0,30,61,117\rangle$ | $\langle 1,22,123,56\rangle$ | $\langle 1,107,119,114\rangle$ |
| | $\langle 0,11,57,91\rangle$ | $\langle 0,60,89,59\rangle$ | $\langle 1,33,42,80\rangle$ | $\langle 0,32,84,109\rangle$ | $\langle 1,23,113,71\rangle$ | $\langle \infty_0,0,19,2\rangle$ |
| | $\langle 0,18,130,4\rangle$ | $\langle 0,62,63,51\rangle$ | $\langle 1,36,80,98\rangle$ | $\langle 0,33,40,120\rangle$ | $\langle 1,57,112,29\rangle$ | $\langle \infty_2,0,81,1\rangle$ |
| | $\langle 0,23,71,21\rangle$ | $\langle 0,8,126,76\rangle$ | $\langle 1,45,50,34\rangle$ | $\langle 0,42,85,125\rangle$ | $\langle 1,59,95,104\rangle$ | $\langle \infty_3,0,93,5\rangle$ |
| | $\langle 0,35,98,47\rangle$ | $\langle 1,12,88,69\rangle$ | $\langle 1,55,82,94\rangle$ | $\langle 0,54,122,94\rangle$ | $\langle 1,83,106,11\rangle$ | $\langle \infty_1,0,37,10\rangle$ |
| | $\langle 0,47,87,71\rangle$ | $\langle 1,14,60,86\rangle$ | $\langle 1,65,68,28\rangle$ | $\langle 0,67,104,16\rangle$ | | |

Table 3: The Codewords of Optimal $(n,5,[3,1])_3$-codes for $n \in \{11,17,23,29,35\}$

| $n$ | Codewords | | | | | |
|---|---|---|---|---|---|---|
| 11 | $\langle 0,1,2,9\rangle$ | $\langle 1,3,5,0\rangle$ | $\langle 2,4,8,1\rangle$ | $\langle 4,5,7,9\rangle$ | $\langle 1,7,9,10\rangle$ | $\langle 4,9,10,2\rangle$ |
| | $\langle 0,5,6,4\rangle$ | $\langle 1,4,6,3\rangle$ | $\langle 2,6,9,5\rangle$ | $\langle 5,8,9,3\rangle$ | $\langle 1,8,10,6\rangle$ | $\langle 6,7,10,0\rangle$ |
| | $\langle 0,7,8,5\rangle$ | $\langle 2,3,7,4\rangle$ | $\langle 3,6,8,7\rangle$ | $\langle 0,3,10,8\rangle$ | $\langle 2,5,10,7\rangle$ | |
| 17 | $\langle 0,1,5,7\rangle$ | $\langle 1,8,14,4\rangle$ | $\langle 5,6,13,4\rangle$ | $\langle 2,5,16,11\rangle$ | $\langle 4,6,14,15\rangle$ | $\langle 9,13,16,2\rangle$ |
| | $\langle 5,7,8,2\rangle$ | $\langle 1,9,10,8\rangle$ | $\langle 0,10,15,4\rangle$ | $\langle 2,6,15,13\rangle$ | $\langle 4,9,11,13\rangle$ | $\langle 0,12,16,13\rangle$ |
| | $\langle 6,7,9,3\rangle$ | $\langle 2,3,12,5\rangle$ | $\langle 0,2,13,14\rangle$ | $\langle 3,14,16,9\rangle$ | $\langle 5,10,14,1\rangle$ | $\langle 11,14,15,0\rangle$ |
| | $\langle 0,3,9,15\rangle$ | $\langle 2,4,7,16\rangle$ | $\langle 0,7,14,12\rangle$ | $\langle 3,4,13,10\rangle$ | $\langle 5,9,12,16\rangle$ | $\langle 7,12,15,10\rangle$ |
| | $\langle 0,4,8,11\rangle$ | $\langle 2,8,10,0\rangle$ | $\langle 1,2,11,10\rangle$ | $\langle 3,5,11,14\rangle$ | $\langle 6,10,16,5\rangle$ | $\langle 8,11,12,15\rangle$ |
| | $\langle 0,6,11,2\rangle$ | $\langle 2,9,14,7\rangle$ | $\langle 1,3,15,16\rangle$ | $\langle 3,7,10,11\rangle$ | $\langle 7,11,16,1\rangle$ | $\langle 10,11,13,16\rangle$ |
| | $\langle 1,6,12,0\rangle$ | $\langle 3,6,8,12\rangle$ | $\langle 1,4,16,12\rangle$ | $\langle 4,10,12,2\rangle$ | $\langle 8,15,16,6\rangle$ | $\langle 12,13,14,11\rangle$ |
| | $\langle 1,7,13,9\rangle$ | $\langle 4,5,15,3\rangle$ | | | | |
| 23 | $\langle 0,3,4,8\rangle$ | $\langle 3,9,13,4\rangle$ | $\langle 2,14,15,1\rangle$ | $\langle 6,12,20,1\rangle$ | $\langle 11,14,18,8\rangle$ | $\langle 6,15,22,13\rangle$ |
| | $\langle 0,5,6,3\rangle$ | $\langle 5,9,17,6\rangle$ | $\langle 2,3,21,12\rangle$ | $\langle 6,14,21,4\rangle$ | $\langle 12,18,19,6\rangle$ | $\langle 7,14,16,18\rangle$ |
| | $\langle 7,8,9,1\rangle$ | $\langle 0,11,16,2\rangle$ | $\langle 3,18,22,5\rangle$ | $\langle 6,7,11,20\rangle$ | $\langle 14,20,22,2\rangle$ | $\langle 8,11,19,12\rangle$ |
| | $\langle 0,1,22,4\rangle$ | $\langle 0,17,18,7\rangle$ | $\langle 3,7,12,16\rangle$ | $\langle 6,8,10,22\rangle$ | $\langle 15,16,19,5\rangle$ | $\langle 8,15,18,21\rangle$ |
| | $\langle 0,7,21,5\rangle$ | $\langle 0,2,19,13\rangle$ | $\langle 4,11,13,0\rangle$ | $\langle 7,17,22,8\rangle$ | $\langle 16,21,22,1\rangle$ | $\langle 9,11,15,14\rangle$ |
| | $\langle 1,3,8,20\rangle$ | $\langle 0,8,20,19\rangle$ | $\langle 4,12,14,5\rangle$ | $\langle 7,19,20,4\rangle$ | $\langle 2,11,20,10\rangle$ | $\langle 9,12,22,17\rangle$ |
| | $\langle 1,4,6,19\rangle$ | $\langle 0,9,10,15\rangle$ | $\langle 4,18,21,3\rangle$ | $\langle 8,13,21,6\rangle$ | $\langle 3,10,14,21\rangle$ | $\langle 9,16,18,20\rangle$ |
| | $\langle 1,7,13,2\rangle$ | $\langle 1,10,16,6\rangle$ | $\langle 4,7,15,12\rangle$ | $\langle 8,16,17,0\rangle$ | $\langle 3,11,17,13\rangle$ | $\langle 10,11,22,19\rangle$ |
| | $\langle 2,4,5,18\rangle$ | $\langle 1,11,12,9\rangle$ | $\langle 4,8,22,11\rangle$ | $\langle 9,20,21,0\rangle$ | $\langle 3,15,20,22\rangle$ | $\langle 10,18,20,14\rangle$ |
| | $\langle 2,6,9,11\rangle$ | $\langle 1,19,21,8\rangle$ | $\langle 4,9,19,21\rangle$ | $\langle 0,12,15,18\rangle$ | $\langle 4,10,17,16\rangle$ | $\langle 12,13,16,15\rangle$ |
| | $\langle 2,7,10,0\rangle$ | $\langle 1,2,18,22\rangle$ | $\langle 5,13,15,8\rangle$ | $\langle 0,13,14,12\rangle$ | $\langle 4,16,20,13\rangle$ | $\langle 12,17,21,19\rangle$ |
| | $\langle 2,8,12,7\rangle$ | $\langle 1,5,20,15\rangle$ | $\langle 5,19,22,0\rangle$ | $\langle 1,15,17,11\rangle$ | $\langle 5,10,12,20\rangle$ | $\langle 13,17,20,18\rangle$ |
| | $\langle 3,5,16,9\rangle$ | $\langle 1,9,14,16\rangle$ | $\langle 5,7,18,11\rangle$ | $\langle 10,13,19,3\rangle$ | $\langle 5,11,21,16\rangle$ | $\langle 14,17,19,22\rangle$ |
| | $\langle 3,6,19,2\rangle$ | $\langle 2,13,22,9\rangle$ | $\langle 5,8,14,17\rangle$ | $\langle 10,15,21,7\rangle$ | $\langle 6,13,18,10\rangle$ | |
| 29 | $\langle 1,3,5,8\rangle$ | $\langle 1,8,14,26\rangle$ | $\langle 6,11,14,8\rangle$ | $\langle 10,11,13,9\rangle$ | $\langle 4,13,28,27\rangle$ | $\langle 9,18,23,27\rangle$ |
| | $\langle 2,3,6,0\rangle$ | $\langle 1,9,25,24\rangle$ | $\langle 6,17,28,3\rangle$ | $\langle 10,16,27,4\rangle$ | $\langle 4,14,20,25\rangle$ | $\langle 10,12,15,21\rangle$ |
| | $\langle 0,1,2,22\rangle$ | $\langle 2,12,18,7\rangle$ | $\langle 6,26,27,7\rangle$ | $\langle 11,16,28,1\rangle$ | $\langle 4,15,17,10\rangle$ | $\langle 10,14,22,24\rangle$ |
| | $\langle 0,4,11,6\rangle$ | $\langle 2,14,21,9\rangle$ | $\langle 6,9,12,26\rangle$ | $\langle 11,17,20,0\rangle$ | $\langle 4,16,19,15\rangle$ | $\langle 10,17,24,25\rangle$ |
| | $\langle 0,6,7,12\rangle$ | $\langle 2,20,28,5\rangle$ | $\langle 7,17,19,8\rangle$ | $\langle 12,13,19,4\rangle$ | $\langle 4,22,23,18\rangle$ | $\langle 10,25,28,15\rangle$ |
| | $\langle 0,9,26,8\rangle$ | $\langle 2,4,25,21\rangle$ | $\langle 7,23,28,9\rangle$ | $\langle 12,22,25,1\rangle$ | $\langle 5,10,18,22\rangle$ | $\langle 11,15,18,28\rangle$ |
| | $\langle 1,4,6,14\rangle$ | $\langle 2,5,24,18\rangle$ | $\langle 8,15,24,7\rangle$ | $\langle 13,15,16,5\rangle$ | $\langle 5,11,26,15\rangle$ | $\langle 11,19,23,22\rangle$ |
| | $\langle 2,8,9,12\rangle$ | $\langle 3,19,21,7\rangle$ | $\langle 8,18,28,6\rangle$ | $\langle 13,23,26,0\rangle$ | $\langle 5,21,28,19\rangle$ | $\langle 11,21,25,17\rangle$ |
| | $\langle 3,4,10,1\rangle$ | $\langle 3,24,28,4\rangle$ | $\langle 8,19,25,0\rangle$ | $\langle 14,15,25,2\rangle$ | $\langle 6,10,20,13\rangle$ | $\langle 12,14,28,23\rangle$ |
| | $\langle 3,7,11,2\rangle$ | $\langle 3,25,26,9\rangle$ | $\langle 9,11,24,5\rangle$ | $\langle 14,18,24,1\rangle$ | $\langle 6,15,22,25\rangle$ | $\langle 12,17,27,14\rangle$ |
| | $\langle 4,5,12,2\rangle$ | $\langle 3,8,12,15\rangle$ | $\langle 9,16,22,2\rangle$ | $\langle 16,23,24,3\rangle$ | $\langle 6,16,21,24\rangle$ | $\langle 12,20,23,17\rangle$ |
| | $\langle 6,8,13,2\rangle$ | $\langle 3,9,15,20\rangle$ | $\langle 9,27,28,0\rangle$ | $\langle 17,21,22,4\rangle$ | $\langle 6,18,19,11\rangle$ | $\langle 12,21,26,28\rangle$ |
| | $\langle 7,8,10,3\rangle$ | $\langle 4,24,27,9\rangle$ | $\langle 0,10,21,16\rangle$ | $\langle 19,24,26,2\rangle$ | $\langle 6,24,25,20\rangle$ | $\langle 13,14,27,19\rangle$ |
| | $\langle 7,9,14,6\rangle$ | $\langle 4,7,18,13\rangle$ | $\langle 0,12,24,13\rangle$ | $\langle 2,10,26,11\rangle$ | $\langle 7,12,16,25\rangle$ | $\langle 13,18,25,23\rangle$ |
| | $\langle 0,22,27,3\rangle$ | $\langle 4,8,26,22\rangle$ | $\langle 0,14,23,21\rangle$ | $\langle 2,11,27,13\rangle$ | $\langle 7,15,27,17\rangle$ | $\langle 13,22,24,11\rangle$ |
| | $\langle 0,3,13,25\rangle$ | $\langle 4,9,21,23\rangle$ | $\langle 0,16,25,14\rangle$ | $\langle 2,15,23,24\rangle$ | $\langle 7,20,25,18\rangle$ | $\langle 14,16,26,18\rangle$ |
| | $\langle 0,5,15,23\rangle$ | $\langle 5,14,19,3\rangle$ | $\langle 0,18,20,15\rangle$ | $\langle 2,16,17,19\rangle$ | $\langle 7,22,26,20\rangle$ | $\langle 15,19,20,26\rangle$ |
| | $\langle 0,8,17,27\rangle$ | $\langle 5,20,22,9\rangle$ | $\langle 0,19,28,24\rangle$ | $\langle 2,19,22,10\rangle$ | $\langle 8,11,22,16\rangle$ | $\langle 15,26,28,16\rangle$ |
| | $\langle 1,10,23,6\rangle$ | $\langle 5,25,27,6\rangle$ | $\langle 1,11,12,10\rangle$ | $\langle 3,14,17,13\rangle$ | $\langle 8,20,27,24\rangle$ | $\langle 17,18,26,24\rangle$ |
| | $\langle 1,16,18,0\rangle$ | $\langle 5,6,23,28\rangle$ | $\langle 1,13,17,15\rangle$ | $\langle 3,16,20,10\rangle$ | $\langle 8,21,23,10\rangle$ | $\langle 17,23,25,16\rangle$ |
| | $\langle 1,20,26,4\rangle$ | $\langle 5,7,13,10\rangle$ | $\langle 1,15,21,27\rangle$ | $\langle 3,18,22,26\rangle$ | $\langle 9,10,19,17\rangle$ | $\langle 18,21,27,20\rangle$ |
| | $\langle 1,22,28,7\rangle$ | $\langle 5,8,16,20\rangle$ | $\langle 1,19,27,16\rangle$ | $\langle 3,23,27,12\rangle$ | $\langle 9,13,20,16\rangle$ | $\langle 20,21,24,22\rangle$ |
| | $\langle 1,7,24,21\rangle$ | $\langle 5,9,17,21\rangle$ | | | | |

| $t$ | | | | | | |
|---|---|---|---|---|---|---|
| 35 | ⟨4, 5, 6, 1⟩ | ⟨3, 12, 30, 0⟩ | ⟨0, 14, 25, 11⟩ | ⟨2, 16, 17, 30⟩ | ⟨7, 16, 24, 22⟩ | ⟨12, 21, 26, 24⟩ |
| | ⟨0, 3, 17, 7⟩ | ⟨3, 19, 25, 5⟩ | ⟨0, 16, 29, 25⟩ | ⟨2, 23, 25, 16⟩ | ⟨7, 17, 30, 14⟩ | ⟨13, 15, 29, 30⟩ |
| | ⟨0, 5, 13, 8⟩ | ⟨3, 6, 10, 15⟩ | ⟨0, 18, 21, 14⟩ | ⟨2, 32, 33, 29⟩ | ⟨7, 19, 32, 25⟩ | ⟨13, 16, 30, 20⟩ |
| | ⟨0, 7, 8, 12⟩ | ⟨3, 7, 11, 29⟩ | ⟨0, 19, 23, 28⟩ | ⟨22, 25, 34, 1⟩ | ⟨7, 20, 28, 10⟩ | ⟨13, 17, 23, 21⟩ |
| | ⟨1, 3, 5, 23⟩ | ⟨3, 8, 33, 14⟩ | ⟨0, 20, 32, 19⟩ | ⟨22, 27, 30, 7⟩ | ⟨7, 22, 33, 17⟩ | ⟨13, 18, 34, 33⟩ |
| | ⟨1, 6, 28, 0⟩ | ⟨4, 12, 25, 3⟩ | ⟨0, 28, 31, 20⟩ | ⟨23, 28, 33, 7⟩ | ⟨7, 23, 27, 20⟩ | ⟨13, 20, 31, 24⟩ |
| | ⟨1, 7, 9, 15⟩ | ⟨4, 19, 30, 8⟩ | ⟨1, 10, 20, 34⟩ | ⟨29, 33, 34, 3⟩ | ⟨7, 31, 34, 16⟩ | ⟨13, 25, 32, 23⟩ |
| | ⟨2, 3, 34, 8⟩ | ⟨4, 21, 34, 7⟩ | ⟨1, 12, 14, 27⟩ | ⟨3, 13, 27, 12⟩ | ⟨8, 10, 22, 28⟩ | ⟨14, 15, 27, 26⟩ |
| | ⟨3, 4, 18, 6⟩ | ⟨4, 8, 28, 18⟩ | ⟨1, 13, 24, 14⟩ | ⟨3, 14, 32, 28⟩ | ⟨8, 12, 34, 29⟩ | ⟨14, 16, 31, 13⟩ |
| | ⟨3, 9, 20, 1⟩ | ⟨4, 9, 26, 32⟩ | ⟨1, 16, 18, 21⟩ | ⟨3, 15, 16, 34⟩ | ⟨8, 14, 18, 20⟩ | ⟨14, 17, 34, 24⟩ |
| | ⟨4, 7, 13, 0⟩ | ⟨5, 10, 27, 9⟩ | ⟨1, 21, 23, 22⟩ | ⟨3, 21, 29, 27⟩ | ⟨8, 15, 25, 24⟩ | ⟨14, 20, 22, 31⟩ |
| | ⟨8, 9, 11, 5⟩ | ⟨5, 21, 24, 4⟩ | ⟨1, 22, 32, 24⟩ | ⟨3, 22, 31, 10⟩ | ⟨8, 20, 27, 17⟩ | ⟨14, 26, 28, 34⟩ |
| | ⟨0, 1, 34, 13⟩ | ⟨5, 7, 25, 31⟩ | ⟨10, 11, 34, 0⟩ | ⟨3, 23, 26, 30⟩ | ⟨8, 21, 31, 25⟩ | ⟨15, 24, 32, 16⟩ |
| | ⟨0, 15, 22, 3⟩ | ⟨5, 8, 32, 22⟩ | ⟨10, 12, 31, 8⟩ | ⟨3, 24, 28, 33⟩ | ⟨9, 10, 23, 31⟩ | ⟨15, 26, 30, 10⟩ |
| | ⟨0, 2, 26, 23⟩ | ⟨5, 9, 29, 12⟩ | ⟨10, 16, 25, 7⟩ | ⟨4, 11, 14, 22⟩ | ⟨9, 12, 22, 16⟩ | ⟨16, 20, 21, 23⟩ |
| | ⟨0, 24, 33, 5⟩ | ⟨6, 11, 32, 3⟩ | ⟨10, 17, 21, 3⟩ | ⟨4, 15, 20, 25⟩ | ⟨9, 13, 19, 27⟩ | ⟨17, 18, 22, 27⟩ |
| | ⟨0, 4, 10, 30⟩ | ⟨6, 13, 22, 5⟩ | ⟨11, 12, 15, 1⟩ | ⟨4, 16, 22, 15⟩ | ⟨9, 15, 17, 19⟩ | ⟨17, 20, 29, 16⟩ |
| | ⟨0, 6, 12, 10⟩ | ⟨6, 20, 30, 9⟩ | ⟨11, 16, 28, 4⟩ | ⟨4, 17, 33, 26⟩ | ⟨9, 25, 28, 22⟩ | ⟨17, 26, 32, 31⟩ |
| | ⟨0, 9, 30, 33⟩ | ⟨6, 24, 31, 7⟩ | ⟨11, 18, 30, 2⟩ | ⟨4, 23, 32, 11⟩ | ⟨10, 13, 14, 17⟩ | ⟨17, 27, 28, 13⟩ |
| | ⟨1, 11, 17, 6⟩ | ⟨6, 7, 15, 28⟩ | ⟨11, 23, 24, 9⟩ | ⟨4, 24, 27, 28⟩ | ⟨10, 15, 19, 23⟩ | ⟨18, 23, 31, 29⟩ |
| | ⟨1, 15, 31, 4⟩ | ⟨6, 8, 17, 33⟩ | ⟨14, 19, 24, 2⟩ | ⟨5, 11, 33, 16⟩ | ⟨10, 18, 33, 25⟩ | ⟨18, 28, 29, 19⟩ |
| | ⟨1, 2, 30, 25⟩ | ⟨6, 9, 14, 21⟩ | ⟨14, 23, 29, 8⟩ | ⟨5, 12, 28, 11⟩ | ⟨10, 24, 26, 20⟩ | ⟨19, 21, 27, 10⟩ |
| | ⟨1, 25, 33, 9⟩ | ⟨7, 12, 29, 6⟩ | ⟨14, 30, 33, 4⟩ | ⟨5, 15, 23, 32⟩ | ⟨10, 28, 30, 32⟩ | ⟨19, 22, 28, 30⟩ |
| | ⟨1, 26, 27, 3⟩ | ⟨7, 14, 21, 5⟩ | ⟨15, 21, 33, 8⟩ | ⟨5, 16, 34, 26⟩ | ⟨10, 29, 32, 18⟩ | ⟨19, 31, 33, 22⟩ |
| | ⟨1, 4, 29, 10⟩ | ⟨7, 18, 26, 8⟩ | ⟨15, 28, 34, 2⟩ | ⟨5, 17, 19, 29⟩ | ⟨11, 13, 21, 19⟩ | ⟨20, 24, 34, 18⟩ |
| | ⟨1, 8, 19, 32⟩ | ⟨8, 13, 26, 4⟩ | ⟨16, 27, 32, 1⟩ | ⟨5, 18, 20, 15⟩ | ⟨11, 19, 29, 15⟩ | ⟨20, 26, 33, 28⟩ |
| | ⟨2, 19, 20, 3⟩ | ⟨8, 16, 23, 0⟩ | ⟨17, 25, 31, 2⟩ | ⟨5, 22, 26, 21⟩ | ⟨11, 20, 25, 33⟩ | ⟨21, 28, 32, 12⟩ |
| | ⟨2, 4, 31, 12⟩ | ⟨8, 29, 30, 1⟩ | ⟨18, 25, 27, 0⟩ | ⟨5, 30, 31, 28⟩ | ⟨11, 26, 31, 18⟩ | ⟨22, 24, 29, 23⟩ |
| | ⟨2, 5, 14, 10⟩ | ⟨9, 16, 33, 6⟩ | ⟨19, 26, 34, 9⟩ | ⟨6, 16, 26, 14⟩ | ⟨12, 13, 33, 15⟩ | ⟨24, 25, 30, 27⟩ |
| | ⟨2, 6, 29, 24⟩ | ⟨9, 18, 24, 3⟩ | ⟨2, 11, 22, 13⟩ | ⟨6, 18, 19, 26⟩ | ⟨12, 16, 19, 17⟩ | ⟨25, 26, 29, 13⟩ |
| | ⟨2, 7, 10, 26⟩ | ⟨9, 27, 34, 4⟩ | ⟨2, 12, 27, 18⟩ | ⟨6, 21, 25, 30⟩ | ⟨12, 17, 24, 25⟩ | ⟨27, 29, 31, 34⟩ |
| | ⟨2, 8, 24, 11⟩ | ⟨9, 31, 32, 0⟩ | ⟨2, 13, 28, 31⟩ | ⟨6, 23, 34, 25⟩ | ⟨12, 18, 32, 34⟩ | ⟨30, 32, 34, 17⟩ |
| | ⟨2, 9, 21, 28⟩ | ⟨0, 11, 27, 31⟩ | ⟨2, 15, 18, 22⟩ | ⟨6, 27, 33, 32⟩ | ⟨12, 20, 23, 26⟩ | |

Table 4: Base Codewords of Optimal $(6t + 5, 5, [3, 1])_3$-codes for $t \in \{6, 8, 10, 12, 14, 16\}$

| $t$ | | | Base Codewords | | | |
|---|---|---|---|---|---|---|
| 6 | ⟨3, 9, 14, 8⟩ | ⟨15, 17, 27, 4⟩ | ⟨0, 24, $\infty_8$, 4⟩ | ⟨0, 26, $\infty_0$, 27⟩ | ⟨6, 21, $\infty_3$, 29⟩ | ⟨6, 29, $\infty_{10}$, 4⟩ |
| | ⟨0, 2, 14, 20⟩ | ⟨5, 15, 26, 29⟩ | ⟨8, 26, $\infty_4$, 4⟩ | ⟨6, 18, $\infty_1$, 14⟩ | ⟨6, 22, $\infty_5$, 20⟩ | ⟨7, 24, $\infty_6$, 10⟩ |
| | ⟨6, 14, 28, 1⟩ | ⟨15, 16, 22, 13⟩ | ⟨0, 19, $\infty_2$, 15⟩ | ⟨6, 19, $\infty_9$, 28⟩ | ⟨6, 26, $\infty_7$, 16⟩ | |
| 8 | ⟨0, 4, 9, 2⟩ | ⟨1, 26, 29, 22⟩ | ⟨21, 22, 31, 1⟩ | ⟨0, 35, $\infty_0$, 4⟩ | ⟨0, 36, $\infty_6$, 20⟩ | ⟨0, 41, $\infty_4$, 28⟩ |
| | ⟨0, 7, 15, 3⟩ | ⟨1, 28, 32, 13⟩ | ⟨21, 23, 29, 15⟩ | ⟨0, 22, $\infty_7$, 41⟩ | ⟨0, 37, $\infty_1$, 34⟩ | ⟨1, 40, $\infty_3$, 16⟩ |
| | ⟨0, 2, 20, 10⟩ | ⟨1, 30, 35, 17⟩ | ⟨0, 23, $\infty_8$, 1⟩ | ⟨0, 24, $\infty_9$, 13⟩ | ⟨0, 38, $\infty_2$, 27⟩ | ⟨0, 26, $\infty_{10}$, 30⟩ |
| | ⟨0, 11, 32, 22⟩ | ⟨1, 34, 41, 40⟩ | ⟨0, 30, $\infty_5$, 5⟩ | | | |
| 10 | ⟨5, 6, 8, 31⟩ | ⟨5, 10, 14, 20⟩ | ⟨14, 25, 45, 15⟩ | ⟨29, 33, 39, 36⟩ | ⟨12, 33, $\infty_3$, 50⟩ | ⟨12, 47, $\infty_2$, 23⟩ |
| | ⟨0, 32, 50, 31⟩ | ⟨5, 11, 25, 14⟩ | ⟨15, 28, 52, 10⟩ | ⟨11, 49, $\infty_4$, 2⟩ | ⟨12, 38, $\infty_6$, 25⟩ | ⟨12, 48, $\infty_5$, 16⟩ |
| | ⟨15, 34, 42, 9⟩ | ⟨13, 27, 38, 42⟩ | ⟨27, 47, 52, 26⟩ | ⟨12, 40, $\infty_0$, 4⟩ | ⟨12, 42, $\infty_7$, 51⟩ | ⟨13, 37, $\infty_8$, 53⟩ |
| | ⟨2, 14, 31, 44⟩ | ⟨14, 22, 29, 21⟩ | ⟨29, 30, 42, 52⟩ | ⟨12, 30, $\infty_9$, 32⟩ | ⟨12, 45, $\infty_1$, 46⟩ | ⟨11, 27, $\infty_{10}$, 25⟩ |
| | ⟨4, 14, 53, 16⟩ | | | | | |

| t | | | | | | |
|---|---|---|---|---|---|---|
| 12 | $\langle 0,1,3,57 \rangle$ | $\langle 1,23,56,6 \rangle$ | $\langle 1,46,52,22 \rangle$ | $\langle 33,34,47,32 \rangle$ | $\langle 0,46,\infty_8,44 \rangle$ | $\langle 0,65,\infty_9,13 \rangle$ |
| | $\langle 0,4,9,47 \rangle$ | $\langle 2,45,50,5 \rangle$ | $\langle 1,48,58,54 \rangle$ | $\langle 33,40,58,34 \rangle$ | $\langle 0,52,\infty_5,19 \rangle$ | $\langle 1,63,\infty_1,33 \rangle$ |
| | $\langle 0,6,13,7 \rangle$ | $\langle 0,15,25,60 \rangle$ | $\langle 1,50,59,18 \rangle$ | $\langle 33,45,49,44 \rangle$ | $\langle 0,59,\infty_6,33 \rangle$ | $\langle 1,64,\infty_2,43 \rangle$ |
| | $\langle 0,17,54,8 \rangle$ | $\langle 1,22,57,15 \rangle$ | $\langle 2,43,46,61 \rangle$ | $\langle 1,40,\infty_4,3 \rangle$ | $\langle 0,60,\infty_0,64 \rangle$ | $\langle 0,40,\infty_{10},50 \rangle$ |
| | $\langle 0,19,53,4 \rangle$ | $\langle 1,37,39,16 \rangle$ | $\langle 2,52,63,24 \rangle$ | $\langle 0,42,\infty_7,58 \rangle$ | $\langle 0,64,\infty_3,61 \rangle$ | |
| 14 | $\langle 1,2,5,55 \rangle$ | $\langle 11,39,69,6 \rangle$ | $\langle 9,60,66,76 \rangle$ | $\langle 11,58,61,63 \rangle$ | $\langle 5,77,\infty_2,6 \rangle$ | $\langle 4,74,\infty_0,73 \rangle$ |
| | $\langle 1,13,29,5 \rangle$ | $\langle 39,58,74,0 \rangle$ | $\langle 9,64,75,57 \rangle$ | $\langle 11,60,65,44 \rangle$ | $\langle 4,39,\infty_8,17 \rangle$ | $\langle 4,75,\infty_1,65 \rangle$ |
| | $\langle 1,8,34,15 \rangle$ | $\langle 4,19,45,25 \rangle$ | $\langle 9,65,77,34 \rangle$ | $\langle 39,41,56,12 \rangle$ | $\langle 4,40,\infty_4,68 \rangle$ | $\langle 5,39,\infty_3,51 \rangle$ |
| | $\langle 8,10,52,4 \rangle$ | $\langle 4,21,44,61 \rangle$ | $\langle 10,63,70,39 \rangle$ | $\langle 40,48,61,10 \rangle$ | $\langle 4,52,\infty_5,66 \rangle$ | $\langle 5,42,\infty_9,14 \rangle$ |
| | $\langle 1,10,15,75 \rangle$ | $\langle 5,24,48,27 \rangle$ | $\langle 11,41,70,67 \rangle$ | $\langle 4,56,\infty_6,3 \rangle$ | $\langle 4,65,\infty_7,63 \rangle$ | $\langle 4,42,\infty_{10},16 \rangle$ |
| | $\langle 1,11,32,30 \rangle$ | $\langle 5,44,69,33 \rangle$ | $\langle 11,56,57,48 \rangle$ | | | |
| 16 | $\langle 0,2,3,47 \rangle$ | $\langle 0,32,39,34 \rangle$ | $\langle 2,10,71,74 \rangle$ | $\langle 2,70,85,81 \rangle$ | $\langle 45,64,68,22 \rangle$ | $\langle 1,83,\infty_1,50 \rangle$ |
| | $\langle 0,5,9,83 \rangle$ | $\langle 1,21,50,32 \rangle$ | $\langle 2,55,58,83 \rangle$ | $\langle 2,80,89,73 \rangle$ | $\langle 1,81,\infty_0,0 \rangle$ | $\langle 1,85,\infty_2,71 \rangle$ |
| | $\langle 2,69,79,8 \rangle$ | $\langle 1,22,48,87 \rangle$ | $\langle 2,57,62,89 \rangle$ | $\langle 3,60,66,55 \rangle$ | $\langle 1,45,\infty_6,18 \rangle$ | $\langle 1,86,\infty_3,33 \rangle$ |
| | $\langle 3,73,84,6 \rangle$ | $\langle 1,27,73,28 \rangle$ | $\langle 2,61,68,75 \rangle$ | $\langle 3,67,79,23 \rangle$ | $\langle 1,49,\infty_7,24 \rangle$ | $\langle 1,87,\infty_4,30 \rangle$ |
| | $\langle 0,14,29,19 \rangle$ | $\langle 1,46,74,43 \rangle$ | $\langle 2,64,77,69 \rangle$ | $\langle 45,47,65,6 \rangle$ | $\langle 1,51,\infty_8,49 \rangle$ | $\langle 1,89,\infty_5,14 \rangle$ |
| | $\langle 0,17,35,26 \rangle$ | $\langle 1,52,53,70 \rangle$ | $\langle 2,67,81,58 \rangle$ | $\langle 45,61,69,7 \rangle$ | $\langle 1,55,\infty_9,42 \rangle$ | $\langle 1,59,\infty_{10},44 \rangle$ |
| | $\langle 0,22,34,85 \rangle$ | | | | | |

Table 5: Base Codewords of Optimal $(6t+5,5,[3,1])_3$-codes for $t \in \{7,9,11,13,15\}$

| t | Base Codewords | | | | | |
|---|---|---|---|---|---|---|
| 7 | $\langle 3,4,6,1 \rangle$ | $\langle 3,19,31,7 \rangle$ | $\langle 0,26,\infty_8,6 \rangle$ | $\langle 0,20,\infty_5,21 \rangle$ | $\langle 0,23,\infty_7,12 \rangle$ | $\langle 7,34,\infty_3,15 \rangle$ |
| | $\langle 0,4,12,5 \rangle$ | $\langle 0,25,29,18 \rangle$ | $\langle 0,9,\infty_0,19 \rangle$ | $\langle 0,21,\infty_6,17 \rangle$ | $\langle 18,27,\infty_0,1 \rangle$ | $\langle 10,25,\infty_1,24 \rangle$ |
| | $\langle 0,5,18,30 \rangle$ | $\langle 19,20,27,6 \rangle$ | $\langle 7,21,\infty_4,3 \rangle$ | $\langle 0,22,\infty_9,20 \rangle$ | $\langle 7,24,\infty_2,29 \rangle$ | $\langle 12,18,\infty_{10},20 \rangle$ |
| | $\langle 0,7,19,23 \rangle$ | $\langle 18,20,23,26 \rangle$ | | | | |
| 9 | $\langle 0,1,3,41 \rangle$ | $\langle 4,10,45,2 \rangle$ | $\langle 22,24,25,28 \rangle$ | $\langle 0,33,\infty_5,9 \rangle$ | $\langle 0,43,\infty_1,33 \rangle$ | $\langle 11,32,\infty_9,12 \rangle$ |
| | $\langle 0,8,24,14 \rangle$ | $\langle 5,14,36,8 \rangle$ | $\langle 24,37,44,30 \rangle$ | $\langle 0,37,\infty_7,5 \rangle$ | $\langle 24,36,\infty_0,1 \rangle$ | $\langle 11,41,\infty_4,15 \rangle$ |
| | $\langle 2,6,19,30 \rangle$ | $\langle 0,10,25,46 \rangle$ | $\langle 25,27,33,12 \rangle$ | $\langle 0,38,\infty_6,8 \rangle$ | $\langle 0,36,\infty_{10},11 \rangle$ | $\langle 11,45,\infty_3,43 \rangle$ |
| | $\langle 3,8,26,10 \rangle$ | $\langle 0,29,32,20 \rangle$ | $\langle 31,41,46,42 \rangle$ | $\langle 0,12,\infty_0,25 \rangle$ | $\langle 10,38,\infty_8,20 \rangle$ | $\langle 12,32,\infty_2,39 \rangle$ |
| 11 | $\langle 0,4,9,8 \rangle$ | $\langle 24,56,57,7 \rangle$ | $\langle 11,32,52,49 \rangle$ | $\langle 31,40,54,15 \rangle$ | $\langle 30,45,\infty_0,1 \rangle$ | $\langle 5,59,\infty_1,21 \rangle$ |
| | $\langle 2,8,15,5 \rangle$ | $\langle 4,18,35,30 \rangle$ | $\langle 13,24,39,58 \rangle$ | $\langle 0,35,\infty_2,1 \rangle$ | $\langle 4,57,\infty_7,26 \rangle$ | $\langle 13,53,\infty_3,37 \rangle$ |
| | $\langle 6,8,9,36 \rangle$ | $\langle 5,30,57,12 \rangle$ | $\langle 16,34,53,35 \rangle$ | $\langle 0,15,\infty_0,31 \rangle$ | $\langle 5,33,\infty_5,11 \rangle$ | $\langle 16,54,\infty_4,52 \rangle$ |
| | $\langle 0,39,44,12 \rangle$ | $\langle 6,49,55,58 \rangle$ | $\langle 18,52,54,28 \rangle$ | $\langle 14,41,\infty_9,5 \rangle$ | $\langle 5,35,\infty_6,40 \rangle$ | $\langle 14,30,\infty_{10},37 \rangle$ |
| | $\langle 12,41,54,1 \rangle$ | $\langle 7,17,25,54 \rangle$ | $\langle 30,34,42,36 \rangle$ | $\langle 3,53,\infty_8,46 \rangle$ | | |
| 13 | $\langle 3,4,6,50 \rangle$ | $\langle 1,13,27,70 \rangle$ | $\langle 4,38,57,69 \rangle$ | $\langle 20,42,58,43 \rangle$ | $\langle 2,68,\infty_4,21 \rangle$ | $\langle 7,67,\infty_1,28 \rangle$ |
| | $\langle 0,39,43,2 \rangle$ | $\langle 1,16,37,51 \rangle$ | $\langle 7,58,68,71 \rangle$ | $\langle 20,49,52,45 \rangle$ | $\langle 3,70,\infty_5,25 \rangle$ | $\langle 17,69,\infty_3,20 \rangle$ |
| | $\langle 1,6,10,15 \rangle$ | $\langle 1,18,26,14 \rangle$ | $\langle 12,60,67,50 \rangle$ | $\langle 37,49,62,33 \rangle$ | $\langle 36,54,\infty_0,1 \rangle$ | $\langle 18,53,\infty_9,59 \rangle$ |
| | $\langle 0,13,20,40 \rangle$ | $\langle 13,55,60,4 \rangle$ | $\langle 12,62,71,43 \rangle$ | $\langle 5,59,\infty_8,4 \rangle$ | $\langle 6,43,\infty_6,21 \rangle$ | $\langle 23,68,\infty_7,24 \rangle$ |
| | $\langle 0,30,56,25 \rangle$ | $\langle 15,43,64,7 \rangle$ | $\langle 14,58,60,62 \rangle$ | $\langle 0,18,\infty_0,37 \rangle$ | $\langle 6,69,\infty_2,10 \rangle$ | $\langle 9,42,\infty_{10},51 \rangle$ |
| | $\langle 0,40,41,16 \rangle$ | $\langle 39,61,69,7 \rangle$ | | | | |
| 15 | $\langle 0,5,11,1 \rangle$ | $\langle 0,60,73,78 \rangle$ | $\langle 4,32,74,27 \rangle$ | $\langle 25,44,80,70 \rangle$ | $\langle 0,21,\infty_0,43 \rangle$ | $\langle 42,63,\infty_0,1 \rangle$ |
| | $\langle 2,3,5,61 \rangle$ | $\langle 1,10,14,61 \rangle$ | $\langle 7,27,52,16 \rangle$ | $\langle 25,48,58,78 \rangle$ | $\langle 15,49,\infty_1,6 \rangle$ | $\langle 6,46,\infty_5,63 \rangle$ |
| | $\langle 0,53,58,8 \rangle$ | $\langle 10,56,58,4 \rangle$ | $\langle 7,58,61,55 \rangle$ | $\langle 30,57,73,68 \rangle$ | $\langle 3,42,\infty_8,82 \rangle$ | $\langle 15,67,\infty_9,56 \rangle$ |
| | $\langle 2,46,73,9 \rangle$ | $\langle 2,49,74,73 \rangle$ | $\langle 8,23,40,42 \rangle$ | $\langle 34,49,56,18 \rangle$ | $\langle 3,44,\infty_6,19 \rangle$ | $\langle 16,53,\infty_3,82 \rangle$ |
| | $\langle 0,12,19,42 \rangle$ | $\langle 25,42,51,7 \rangle$ | $\langle 8,70,82,11 \rangle$ | $\langle 42,50,61,46 \rangle$ | $\langle 3,69,\infty_7,21 \rangle$ | $\langle 18,54,\infty_2,68 \rangle$ |
| | $\langle 0,49,50,62 \rangle$ | $\langle 4,22,30,32 \rangle$ | $\langle 9,44,72,79 \rangle$ | $\langle 43,61,65,71 \rangle$ | $\langle 3,83,\infty_4,57 \rangle$ | $\langle 19,75,\infty_{10},52 \rangle$ |

Table 6: Base Codewords of [2, 2]-GDC(5)s of Type $8^u 10^1$

| $u$ | Base Codewords | | | | | |
|---|---|---|---|---|---|---|
| 5 | $\langle 0, 18, 1, 9 \rangle$ | $\langle 1, 19, 10, 23 \rangle$ | $\langle \infty_1, 1, 9, 22 \rangle$ | $\langle 34, 21, 0, \infty_0 \rangle$ | $\langle \infty_0, 0, 29, 32 \rangle$ | $\langle \infty_5, 0, 18, 27 \rangle$ |
|   | $\langle 0, 14, 8, 16 \rangle$ | $\langle 13, 2, 1, \infty_3 \rangle$ | $\langle \infty_8, 1, 2, 25 \rangle$ | $\langle 34, 23, 1, \infty_1 \rangle$ | $\langle \infty_2, 1, 13, 14 \rangle$ | $\langle \infty_6, 0, 26, 37 \rangle$ |
|   | $\langle 0, 37, 4, 11 \rangle$ | $\langle 22, 5, 1, \infty_6 \rangle$ | $\langle 17, 16, 0, \infty_7 \rangle$ | $\langle 35, 28, 1, \infty_4 \rangle$ | $\langle \infty_3, 1, 17, 28 \rangle$ | $\langle \infty_7, 0, 21, 28 \rangle$ |
|   | $\langle 1, 37, 0, 34 \rangle$ | $\langle 24, 3, 1, \infty_5 \rangle$ | $\langle 29, 26, 0, \infty_9 \rangle$ | $\langle 38, 15, 1, \infty_2 \rangle$ | $\langle \infty_4, 1, 18, 35 \rangle$ | $\langle \infty_9, 1, 30, 33 \rangle$ |
|   | $\langle 0, 16, 12, 38 \rangle$ | $\langle 39, 8, 1, \infty_8 \rangle$ | | | | |
| 6 | $\langle 0, 28, 9, 38 \rangle$ | $\langle 0, 19, 22, 45 \rangle$ | $\langle 6, 3, 1, \infty_5 \rangle$ | $\langle \infty_5, 1, 0, 41 \rangle$ | $\langle 28, 15, 0, \infty_2 \rangle$ | $\langle 47, 20, 0, \infty_1 \rangle$ |
|   | $\langle 0, 3, 46, 47 \rangle$ | $\langle 1, 15, 12, 20 \rangle$ | $\langle 46, 9, 0, \infty_0 \rangle$ | $\langle \infty_6, 1, 3, 42 \rangle$ | $\langle 32, 27, 1, \infty_8 \rangle$ | $\langle \infty_0, 0, 13, 16 \rangle$ |
|   | $\langle 0, 8, 15, 31 \rangle$ | $\langle 1, 36, 32, 39 \rangle$ | $\langle \infty_2, 0, 1, 32 \rangle$ | $\langle \infty_7, 1, 8, 33 \rangle$ | $\langle 35, 12, 1, \infty_9 \rangle$ | $\langle \infty_1, 0, 14, 25 \rangle$ |
|   | $\langle 0, 16, 21, 35 \rangle$ | $\langle 1, 41, 28, 30 \rangle$ | $\langle \infty_3, 0, 4, 41 \rangle$ | $\langle 21, 16, 1, \infty_7 \rangle$ | $\langle 40, 33, 0, \infty_4 \rangle$ | $\langle \infty_8, 0, 11, 26 \rangle$ |
|   | $\langle 0, 17, 27, 40 \rangle$ | $\langle 1, 45, 14, 17 \rangle$ | $\langle \infty_4, 1, 5, 10 \rangle$ | $\langle 23, 14, 0, \infty_6 \rangle$ | $\langle 41, 10, 1, \infty_3 \rangle$ | $\langle \infty_9, 1, 22, 35 \rangle$ |
| 7 | $\langle 1, 5, 6, 7 \rangle$ | $\langle 0, 4, 43, 45 \rangle$ | $\langle 1, 17, 37, 42 \rangle$ | $\langle 41, 32, 0, \infty_8 \rangle$ | $\langle 54, 39, 0, \infty_2 \rangle$ | $\langle \infty_5, 0, 25, 40 \rangle$ |
|   | $\langle 0, 6, 5, 16 \rangle$ | $\langle 1, 33, 0, 25 \rangle$ | $\langle 1, 27, 23, 39 \rangle$ | $\langle 47, 30, 0, \infty_7 \rangle$ | $\langle \infty_0, 1, 48, 51 \rangle$ | $\langle \infty_6, 1, 32, 55 \rangle$ |
|   | $\langle 0, 1, 12, 38 \rangle$ | $\langle 0, 20, 31, 53 \rangle$ | $\langle 23, 8, 0, \infty_3 \rangle$ | $\langle 48, 11, 1, \infty_6 \rangle$ | $\langle \infty_1, 1, 28, 41 \rangle$ | $\langle \infty_7, 0, 27, 30 \rangle$ |
|   | $\langle 0, 11, 6, 50 \rangle$ | $\langle 0, 31, 18, 34 \rangle$ | $\langle 27, 4, 0, \infty_5 \rangle$ | $\langle 48, 37, 0, \infty_4 \rangle$ | $\langle \infty_2, 0, 19, 46 \rangle$ | $\langle \infty_8, 1, 17, 54 \rangle$ |
|   | $\langle 0, 3, 13, 37 \rangle$ | $\langle 0, 32, 20, 54 \rangle$ | $\langle \infty_4, 0, 4, 51 \rangle$ | $\langle 53, 40, 1, \infty_0 \rangle$ | $\langle \infty_3, 1, 14, 45 \rangle$ | $\langle \infty_9, 0, 15, 32 \rangle$ |
|   | $\langle 0, 39, 1, 47 \rangle$ | $\langle 0, 47, 23, 36 \rangle$ | $\langle 28, 27, 1, \infty_1 \rangle$ | $\langle 54, 31, 1, \infty_9 \rangle$ | | |
| 8 | $\langle 0, 1, 13, 19 \rangle$ | $\langle 1, 60, 2, 31 \rangle$ | $\langle 1, 15, 22, 53 \rangle$ | $\langle \infty_0, 1, 6, 27 \rangle$ | $\langle 31, 22, 0, \infty_8 \rangle$ | $\langle 62, 13, 0, \infty_3 \rangle$ |
|   | $\langle 0, 10, 3, 31 \rangle$ | $\langle 1, 7, 32, 62 \rangle$ | $\langle 1, 26, 23, 46 \rangle$ | $\langle \infty_2, 0, 1, 26 \rangle$ | $\langle 35, 28, 0, \infty_5 \rangle$ | $\langle \infty_1, 1, 35, 54 \rangle$ |
|   | $\langle 0, 62, 9, 12 \rangle$ | $\langle 0, 11, 46, 52 \rangle$ | $\langle 1, 45, 24, 28 \rangle$ | $\langle \infty_6, 1, 4, 11 \rangle$ | $\langle 42, 25, 0, \infty_2 \rangle$ | $\langle \infty_3, 0, 30, 55 \rangle$ |
|   | $\langle 1, 38, 3, 15 \rangle$ | $\langle 0, 45, 43, 58 \rangle$ | $\langle 1, 48, 12, 38 \rangle$ | $\langle \infty_7, 0, 4, 27 \rangle$ | $\langle 50, 23, 1, \infty_1 \rangle$ | $\langle \infty_4, 1, 10, 29 \rangle$ |
|   | $\langle 1, 4, 37, 55 \rangle$ | $\langle 0, 50, 45, 49 \rangle$ | $\langle 14, 5, 0, \infty_6 \rangle$ | $\langle \infty_8, 1, 7, 58 \rangle$ | $\langle 54, 47, 0, \infty_0 \rangle$ | $\langle \infty_5, 0, 47, 60 \rangle$ |
|   | $\langle 1, 46, 5, 16 \rangle$ | $\langle 0, 63, 18, 62 \rangle$ | $\langle 40, 5, 1, \infty_9 \rangle$ | $\langle 26, 15, 0, \infty_7 \rangle$ | $\langle 60, 19, 1, \infty_4 \rangle$ | $\langle \infty_9, 0, 17, 44 \rangle$ |
|   | $\langle 1, 6, 45, 59 \rangle$ | $\langle 1, 14, 21, 51 \rangle$ | | | | |
| 9 | $\langle 0, 3, 5, 51 \rangle$ | $\langle 1, 67, 9, 17 \rangle$ | $\langle 0, 71, 20, 68 \rangle$ | $\langle 1, 71, 43, 66 \rangle$ | $\langle 27, 26, 1, \infty_1 \rangle$ | $\langle \infty_1, 1, 30, 61 \rangle$ |
|   | $\langle 0, 22, 7, 66 \rangle$ | $\langle 0, 10, 24, 32 \rangle$ | $\langle 1, 31, 65, 69 \rangle$ | $\langle 69, 2, 0, \infty_6 \rangle$ | $\langle 30, 23, 0, \infty_7 \rangle$ | $\langle \infty_2, 1, 42, 71 \rangle$ |
|   | $\langle 0, 4, 19, 35 \rangle$ | $\langle 0, 14, 17, 39 \rangle$ | $\langle 1, 47, 18, 33 \rangle$ | $\langle \infty_0, 0, 8, 13 \rangle$ | $\langle 49, 16, 0, \infty_0 \rangle$ | $\langle \infty_3, 1, 27, 58 \rangle$ |
|   | $\langle 0, 48, 2, 37 \rangle$ | $\langle 0, 32, 43, 55 \rangle$ | $\langle 1, 61, 26, 36 \rangle$ | $\langle \infty_7, 0, 4, 21 \rangle$ | $\langle 61, 44, 0, \infty_2 \rangle$ | $\langle \infty_4, 1, 13, 32 \rangle$ |
|   | $\langle 1, 32, 6, 29 \rangle$ | $\langle 0, 47, 53, 60 \rangle$ | $\langle 1, 62, 31, 52 \rangle$ | $\langle \infty_8, 0, 1, 12 \rangle$ | $\langle 62, 33, 0, \infty_9 \rangle$ | $\langle \infty_5, 0, 34, 67 \rangle$ |
|   | $\langle 1, 36, 2, 16 \rangle$ | $\langle 0, 55, 33, 65 \rangle$ | $\langle 1, 63, 57, 62 \rangle$ | $\langle 24, 11, 1, \infty_5 \rangle$ | $\langle 65, 14, 0, \infty_3 \rangle$ | $\langle \infty_6, 0, 64, 71 \rangle$ |
|   | $\langle 1, 57, 5, 25 \rangle$ | $\langle 0, 69, 16, 50 \rangle$ | $\langle 1, 66, 34, 53 \rangle$ | $\langle 24, 13, 0, \infty_4 \rangle$ | $\langle 66, 17, 0, \infty_8 \rangle$ | $\langle \infty_9, 0, 29, 30 \rangle$ |

Table 7: Base Codewords of [2, 2]-GDC(5)s of Type $8^v 14^1$

| $v$ | Base Codewords | | | | | |
|---|---|---|---|---|---|---|
| 5 | $\langle 0, 4, 6, 33 \rangle$ | $\langle 12, 3, 0, \infty_2 \rangle$ | $\langle \infty_5, 1, 4, 17 \rangle$ | $\langle 29, 22, 1, \infty_6 \rangle$ | $\langle \infty_1, 1, 10, 29 \rangle$ | $\langle 32, 29, 0, \infty_{12} \rangle$ |
|   | $\langle 0, 39, 18, 26 \rangle$ | $\langle 23, 2, 0, \infty_8 \rangle$ | $\langle \infty_8, 1, 7, 30 \rangle$ | $\langle 33, 30, 1, \infty_1 \rangle$ | $\langle \infty_3, 1, 25, 34 \rangle$ | $\langle 37, 20, 1, \infty_{13} \rangle$ |
|   | $\langle 5, 4, 1, \infty_5 \rangle$ | $\langle 26, 9, 0, \infty_4 \rangle$ | $\langle \infty_9, 0, 9, 16 \rangle$ | $\langle 39, 10, 1, \infty_0 \rangle$ | $\langle \infty_6, 1, 23, 24 \rangle$ | $\langle \infty_{10}, 1, 14, 35 \rangle$ |
|   | $\langle 8, 1, 0, \infty_9 \rangle$ | $\langle \infty_0, 0, 7, 24 \rangle$ | $\langle 23, 14, 1, \infty_7 \rangle$ | $\langle 39, 28, 0, \infty_3 \rangle$ | $\langle \infty_7, 0, 36, 39 \rangle$ | $\langle \infty_{11}, 1, 22, 33 \rangle$ |
|   | $\langle \infty_2, 0, 1, 4 \rangle$ | $\langle \infty_4, 1, 8, 15 \rangle$ | $\langle 24, 3, 1, \infty_{10} \rangle$ | $\langle \infty_{13}, 0, 3, 34 \rangle$ | $\langle 28, 15, 1, \infty_{11} \rangle$ | $\langle \infty_{12}, 0, 22, 23 \rangle$ |
| 6 | $\langle 0, 40, 2, 47 \rangle$ | $\langle 9, 6, 1, \infty_5 \rangle$ | $\langle \infty_8, 0, 4, 33 \rangle$ | $\langle 41, 16, 0, \infty_7 \rangle$ | $\langle \infty_4, 0, 26, 45 \rangle$ | $\langle 29, 18, 1, \infty_{10} \rangle$ |
|   | $\langle 1, 15, 4, 26 \rangle$ | $\langle 27, 8, 0, \infty_1 \rangle$ | $\langle 32, 21, 1, \infty_3 \rangle$ | $\langle 43, 10, 0, \infty_6 \rangle$ | $\langle \infty_6, 0, 20, 27 \rangle$ | $\langle 35, 20, 1, \infty_{12} \rangle$ |
|   | $\langle 1, 21, 2, 16 \rangle$ | $\langle 46, 5, 1, \infty_0 \rangle$ | $\langle 33, 10, 1, \infty_2 \rangle$ | $\langle 47, 38, 1, \infty_4 \rangle$ | $\langle \infty_7, 1, 23, 34 \rangle$ | $\langle \infty_{10}, 1, 24, 47 \rangle$ |
|   | $\langle 0, 22, 37, 41 \rangle$ | $\langle \infty_0, 1, 0, 33 \rangle$ | $\langle 34, 17, 0, \infty_8 \rangle$ | $\langle \infty_{12}, 1, 5, 42 \rangle$ | $\langle \infty_9, 0, 35, 44 \rangle$ | $\langle \infty_{11}, 1, 36, 39 \rangle$ |
|   | $\langle 1, 27, 18, 46 \rangle$ | $\langle \infty_2, 0, 9, 16 \rangle$ | $\langle 40, 35, 0, \infty_9 \rangle$ | $\langle \infty_1, 1, 27, 28 \rangle$ | $\langle 24, 15, 1, \infty_{11} \rangle$ | $\langle \infty_{13}, 0, 21, 34 \rangle$ |
|   | $\langle 1, 36, 10, 11 \rangle$ | $\langle \infty_5, 0, 5, 28 \rangle$ | $\langle 41, 0, 1, \infty_{13} \rangle$ | $\langle \infty_3, 0, 13, 46 \rangle$ | | |
| 7 | $\langle 0, 24, 4, 6 \rangle$ | $\langle 1, 24, 21, 26 \rangle$ | $\langle \infty_7, 0, 1, 52 \rangle$ | $\langle 49, 48, 1, \infty_2 \rangle$ | $\langle \infty_3, 0, 41, 44 \rangle$ | $\langle 51, 48, 0, \infty_{12} \rangle$ |
| | $\langle 0, 18, 10, 29 \rangle$ | $\langle 1, 31, 41, 47 \rangle$ | $\langle 10, 5, 1, \infty_{11} \rangle$ | $\langle 52, 23, 1, \infty_0 \rangle$ | $\langle \infty_4, 0, 22, 31 \rangle$ | $\langle 54, 39, 1, \infty_{13} \rangle$ |
| | $\langle 0, 20, 33, 43 \rangle$ | $\langle 1, 53, 30, 40 \rangle$ | $\langle 19, 10, 0, \infty_6 \rangle$ | $\langle 53, 22, 0, \infty_9 \rangle$ | $\langle \infty_5, 1, 24, 55 \rangle$ | $\langle \infty_{10}, 0, 39, 54 \rangle$ |
| | $\langle 0, 45, 15, 51 \rangle$ | $\langle 12, 9, 1, \infty_7 \rangle$ | $\langle 25, 20, 1, \infty_5 \rangle$ | $\langle 53, 38, 1, \infty_1 \rangle$ | $\langle \infty_8, 1, 23, 54 \rangle$ | $\langle \infty_{11}, 1, 10, 37 \rangle$ |
| | $\langle 0, 54, 24, 30 \rangle$ | $\langle 36, 5, 0, \infty_4 \rangle$ | $\langle 32, 19, 1, \infty_3 \rangle$ | $\langle \infty_0, 1, 14, 25 \rangle$ | $\langle \infty_9, 1, 48, 51 \rangle$ | $\langle \infty_{12}, 1, 13, 46 \rangle$ |
| | $\langle 0, 55, 16, 40 \rangle$ | $\langle \infty_2, 1, 2, 31 \rangle$ | $\langle 40, 13, 1, \infty_8 \rangle$ | $\langle \infty_1, 0, 18, 55 \rangle$ | $\langle 44, 25, 0, \infty_{10} \rangle$ | $\langle \infty_{13}, 0, 27, 50 \rangle$ |
| | $\langle 1, 17, 16, 28 \rangle$ | $\langle \infty_6, 1, 3, 20 \rangle$ | | | | |
| 8 | $\langle 0, 27, 2, 7 \rangle$ | $\langle 0, 53, 20, 39 \rangle$ | $\langle 40, 5, 1, \infty_2 \rangle$ | $\langle 55, 22, 0, \infty_7 \rangle$ | $\langle \infty_2, 1, 42, 43 \rangle$ | $\langle 18, 13, 1, \infty_{11} \rangle$ |
| | $\langle 1, 53, 3, 8 \rangle$ | $\langle 0, 58, 37, 49 \rangle$ | $\langle 42, 3, 0, \infty_4 \rangle$ | $\langle 56, 39, 1, \infty_6 \rangle$ | $\langle \infty_3, 0, 18, 35 \rangle$ | $\langle 34, 27, 0, \infty_{10} \rangle$ |
| | $\langle 0, 1, 19, 36 \rangle$ | $\langle 1, 24, 29, 55 \rangle$ | $\langle 45, 0, 1, \infty_8 \rangle$ | $\langle 61, 52, 1, \infty_0 \rangle$ | $\langle \infty_4, 0, 28, 59 \rangle$ | $\langle 37, 30, 0, \infty_{13} \rangle$ |
| | $\langle 0, 22, 4, 11 \rangle$ | $\langle 1, 26, 12, 23 \rangle$ | $\langle \infty_6, 1, 7, 48 \rangle$ | $\langle 61, 58, 0, \infty_1 \rangle$ | $\langle \infty_5, 1, 22, 35 \rangle$ | $\langle 54, 51, 0, \infty_{12} \rangle$ |
| | $\langle 0, 15, 14, 60 \rangle$ | $\langle 1, 39, 44, 54 \rangle$ | $\langle 31, 10, 0, \infty_5 \rangle$ | $\langle \infty_0, 1, 39, 60 \rangle$ | $\langle \infty_7, 1, 31, 56 \rangle$ | $\langle \infty_{10}, 0, 26, 63 \rangle$ |
| | $\langle 0, 30, 23, 51 \rangle$ | $\langle 1, 59, 24, 52 \rangle$ | $\langle 38, 29, 1, \infty_3 \rangle$ | $\langle \infty_{13}, 1, 2, 47 \rangle$ | $\langle \infty_8, 0, 41, 44 \rangle$ | $\langle \infty_{11}, 0, 45, 58 \rangle$ |
| | $\langle 0, 35, 29, 52 \rangle$ | $\langle 1, 60, 13, 63 \rangle$ | $\langle 39, 26, 0, \infty_9 \rangle$ | $\langle \infty_1, 0, 12, 15 \rangle$ | $\langle \infty_9, 0, 33, 62 \rangle$ | $\langle \infty_{12}, 1, 11, 50 \rangle$ |
| 9 | $\langle 1, 5, 9, 49 \rangle$ | $\langle 0, 34, 39, 47 \rangle$ | $\langle 1, 61, 32, 71 \rangle$ | $\langle \infty_4, 0, 1, 38 \rangle$ | $\langle 71, 40, 0, \infty_0 \rangle$ | $\langle \infty_9, 0, 46, 61 \rangle$ |
| | $\langle 0, 42, 4, 52 \rangle$ | $\langle 0, 35, 15, 59 \rangle$ | $\langle 1, 68, 12, 56 \rangle$ | $\langle \infty_7, 0, 7, 66 \rangle$ | $\langle \infty_0, 0, 24, 67 \rangle$ | $\langle 48, 27, 1, \infty_{13} \rangle$ |
| | $\langle 0, 48, 6, 40 \rangle$ | $\langle 0, 70, 19, 33 \rangle$ | $\langle 2, 1, 0, \infty_{10} \rangle$ | $\langle 11, 10, 0, \infty_6 \rangle$ | $\langle \infty_2, 1, 41, 42 \rangle$ | $\langle 50, 43, 1, \infty_{11} \rangle$ |
| | $\langle 0, 8, 22, 28 \rangle$ | $\langle 1, 24, 27, 68 \rangle$ | $\langle 23, 8, 1, \infty_8 \rangle$ | $\langle 13, 2, 1, \infty_{12} \rangle$ | $\langle \infty_3, 1, 39, 58 \rangle$ | $\langle \infty_{10}, 1, 22, 29 \rangle$ |
| | $\langle 1, 33, 3, 20 \rangle$ | $\langle 1, 50, 33, 67 \rangle$ | $\langle 25, 4, 0, \infty_5 \rangle$ | $\langle 46, 21, 0, \infty_2 \rangle$ | $\langle \infty_5, 1, 16, 63 \rangle$ | $\langle \infty_{11}, 0, 50, 51 \rangle$ |
| | $\langle 0, 14, 11, 57 \rangle$ | $\langle 1, 51, 54, 57 \rangle$ | $\langle 42, 9, 1, \infty_7 \rangle$ | $\langle 55, 14, 0, \infty_1 \rangle$ | $\langle \infty_6, 1, 59, 66 \rangle$ | $\langle \infty_{12}, 1, 17, 38 \rangle$ |
| | $\langle 0, 19, 41, 53 \rangle$ | $\langle 1, 56, 26, 40 \rangle$ | $\langle 44, 5, 1, \infty_3 \rangle$ | $\langle 67, 60, 0, \infty_9 \rangle$ | $\langle \infty_8, 1, 15, 70 \rangle$ | $\langle \infty_{13}, 1, 21, 36 \rangle$ |
| | $\langle 0, 25, 37, 48 \rangle$ | $\langle 1, 57, 14, 34 \rangle$ | $\langle \infty_1, 0, 8, 49 \rangle$ | $\langle 70, 65, 0, \infty_4 \rangle$ | | |